\documentclass[11pt]{article}
\usepackage[T1]{fontenc}

\usepackage[margin=1in]{geometry}
\usepackage{amsmath, amssymb, amsthm, mathtools}
\usepackage{graphicx}
\usepackage{booktabs}
\usepackage{authblk}
\usepackage{float}
\usepackage{natbib}
\usepackage[colorlinks=true, linkcolor=black, citecolor=black, urlcolor=blue]{hyperref}
\usepackage{xcolor}
\usepackage{caption}
\usepackage{subcaption}
\usepackage{microtype}
\usepackage{tikz}
\usetikzlibrary{arrows.meta,positioning,calc,3d}
\graphicspath{{./}}

\title{Masked complex non-decimated wavelet features for patient-level
       classification of contrast-enhanced mammography}

\author{Sara Antonijevic}
\author{Brani Vidakovic}
\affil{Department of Statistics, Texas A\&M University, College Station, TX}
\date{\today}

\begin{document}
\maketitle

\begin{abstract}
Contrast-enhanced spectral mammography (CESM) acquires two images of each breast, a low-energy image and a recombined contrast image, but two questions central to building a classifier on them remain unsettled: whether the two image types carry comparable malignancy signal, and how a patient's several images should be combined into a single decision. Both are hard to answer reliably, because most published CESM classifiers split cross-validation folds at the image level, letting images of the same patient fall in both training and test sets and inflating reported performance. We pair a masked complex non-decimated wavelet feature bank with an elastic-net logistic classifier, evaluated under repeated patient-grouped nested cross-validation with patient-cluster bootstrap inference on the CDD-CESM dataset ($1{,}880$ images, $308$ patients); under this leakage-free evaluation the inflation from testing on previously seen patients is negligible. On normal-versus-malignant detection, the two acquisitions are statistically
indistinguishable in patient-level AUC under the proposed evaluation framework. Under single-image fusion the contrast image reaches a patient-level AUC of $0.874$ ($95\%$ CI $0.827$--$0.918$) and the low-energy image is statistically indistinguishable from it, yet the two encode malignancy through disjoint, interpretable channels: phase coherence on the low-energy image and magnitude distribution on the contrast image. The framework matches a pretrained ResNet-50 representation at the patient level,
but whereas the frozen deep representation is not directly interpretable at the level
of individual predictors, every predictor in the wavelet representation carries an
explicit physical meaning. The result is a transparent, leakage-free baseline against
which future CESM classifiers can be measured.
\end{abstract}


\section{Introduction}
\label{sec:intro}

Breast cancer remains one of the most commonly diagnosed cancers and
a leading cause of cancer-related mortality among women worldwide
\citep{sung_global_2021, siegel_cancer_2024}, and early detection
through screening substantially improves prognosis. Digital
mammography is the screening standard but exhibits reduced
sensitivity in dense breast tissue \citep{boyd_mammographic_2007,
mandelson_breast_2000}, motivating the development of adjunct
imaging modalities that recover the signal lost to tissue
superposition.

Contrast-Enhanced Spectral Mammography (CESM) is a dual-energy imaging modality that addresses the dense-tissue limitation by acquiring two paired exposures after intravenous administration of an iodinated contrast agent \citep{fallenberg_review,
patel_contrast_2018, sorin_contrast_2018}. The lower-energy
exposure, the dual-energy mammogram (DM), resembles a standard
mammogram and depicts overall breast anatomy; the higher-energy
exposure is sensitive to iodine accumulation, and weighted
subtraction of the two yields a recombined contrast mammogram (CM)
in which normal tissue is suppressed and regions of iodine uptake
stand out. Because malignant tumors develop leaky neovasculature
and take up contrast preferentially, the CM image highlights
cancers that are obscured by overlapping dense tissue on the DM
\citep{Khaled2021Dataset, fallenberg_review,
jochelson_contrast_2017}. From a statistical standpoint, the
consequence of this acquisition design is that each patient
contributes multiple correlated images per examination (two image
types $\times$ two views $\times$ two breasts) and any classifier
operating on CESM must therefore confront within-patient correlation
in both its evaluation framework and its patient-level decision rule.

Existing computer-aided CESM classifiers fall into two broad
families. Deep-learning pipelines, typically convolutional neural
networks fine-tuned from ImageNet-pretrained backbones, report
strong image-level AUCs on CDD-CESM, frequently above $0.90$
\citep{acosta_explainable_2025, mohamed_artificial_2020}, but
produce representations from which no
per-feature physical interpretation is recoverable. Hand-crafted
radiomic pipelines using first- and second-order intensity
statistics, gray-level co-occurrence matrices, and shape descriptors
are more interpretable \citep{losurdo_radiomics_2019,
marino_multiparametric_2020} but typically require lesion-level
region-of-interest delineation, which is labor-intensive at scale
and tightly couples the resulting feature bank to the quality of the
delineation. Wavelet-based representations occupy an intermediate
position \citep{vidakovic_book, ramirez_wavelet_2018,
jeon_breast_2020}, and are the basis for the feature bank developed
in Section~\ref{sec:methods}.

A second gap, orthogonal to the choice of feature representation,
concerns evaluation. Cross-validation schemes that split at the
image level allow correlated images from the same patient to appear
in both training and test folds, and the resulting performance
estimates are systematically biased upward
\citep{rouzrokh_mitigating_2022, varoquaux_machine_2022}. The magnitude of this inflation is setting-dependent and can be large:
\citet{yagis_data_leakage_2021} reported that slice-level cross-validation
raised test-set accuracy by $29$--$55\%$ across four brain MRI datasets relative to subject-level cross-validation. That figure is a worst case from a different modality, in which a single subject contributes many near-identical slices, and it indicates how severe leakage can become rather than predicting its size for CESM; the actual magnitude for a given feature bank is an empirical question, which the patient-grouped design of this paper answers directly (Section~\ref{sec:res-baseline}). CESM is nonetheless a clear instance of the underlying problem, because each patient contributes a median of eight correlated images in the dataset used here, and the issue arises whenever the unit of analysis (the image) is nested within a higher-level unit of inference (the patient). Several recent surveys document that image-level cross-validation remains common in
medical-imaging classification work despite the inflation it produces
\citep{rouzrokh_mitigating_2022, varoquaux_machine_2022}.

This paper makes four contributions. First, we introduce a masked
complex non-decimated wavelet feature bank for multi-image medical
classification, a representation that retains both magnitude and
phase at every scale and restricts all summaries to a multi-scale
eroded anatomical support, eliminating background contamination at
the level of the feature definition rather than the classifier.
Second, we evaluate the feature bank under repeated patient-grouped
nested cross-validation with patient-cluster bootstrap confidence
intervals, providing, to our knowledge, the first patient-grouped leakage-free benchmark for a
wavelet feature family on the CDD-CESM dataset.
Third, we use the framework to answer three methodological questions
about CESM classification that the patient-grouped evaluation makes
tractable: which of the two acquired image types carries the
malignancy signal when the same feature bank is applied to each; how
the choice of patient-level fusion rule affects reported performance;
and whether the modality and fusion findings are sensitive to the
choice of classifier family. Fourth, we benchmark the interpretable
wavelet representation against a pretrained ResNet-50 frozen-feature
representation under identical patient-grouped folds, isolating the
contribution of the feature representation while holding classifier
and evaluation framework fixed.

The remainder of the paper is organized as follows.
Section~\ref{sec:data} introduces the CDD-CESM dataset and its
patient-level structure; Section~\ref{sec:methods} develops the
masked complex NDWT feature representation; Section~\ref{sec:model} describes the elastic-net classifier, the
patient-grouped evaluation framework, and the patient-cluster
bootstrap; Section~\ref{sec:results} reports the empirical findings;
Section~\ref{sec:discussion} positions the contribution in the
broader literature.

\section{Data and patient-level structure}
\label{sec:data}

\paragraph{CDD-CESM cohort.}

We use the public Categorized Digital Database for Low-Energy and Subtracted Contrast-Enhanced Spectral Mammography Images (CDD-CESM)~\citep{Khaled2021Dataset, Khaled2022ScientificData}, hosted by The Cancer Imaging Archive, which comprises $2{,}006$ images from $326$ patients acquired on two different mammography units. The two units are not interchangeable sources of images:
different detectors, calibration, and image-processing chains impart device-specific characteristics to the pixel data, and if the patients imaged on one unit differ systematically from those imaged on the other in case mix or acquisition period, then device identity becomes correlated with the malignancy outcome. A classifier trained on the pooled set could then improve its apparent accuracy by learning to recognize the scanner rather than the disease, a between-scanner confound that would inflate performance estimates in the same way that patient-level leakage does, but at the level of the acquisition device. To eliminate this confound at the source, we restrict the analytic set to the primary device, so that every image in the study is acquired under the same imaging physics and any signal the classifier extracts must reflect tissue rather than scanner. This restriction gives $1{,}880$ images from $308$ patients: $620$ normal, $624$ benign, and $636$ malignant, split evenly across $940$ DM (low-energy) and $940$ CM
(recombined contrast) images. Patients contribute between $2$ and $10$ images each, with a median of $8$ (one DM and one CM image per view, per breast). Each image carries a radiologist-confirmed pathology label together with structured Breast Imaging Reporting and Data System (BI-RADS) metadata.

CDD-CESM provides per-image pathology annotations. For $47$ breasts ($87$ images) on the analytic device the annotation differed across the two acquisitions or the two views of the same breast. In every one of
these cases the disagreement took the same form: an image carried a normal annotation while another acquisition or view of the same breast carried a benign or malignant finding, and the normal reads fell
predominantly on the recombined contrast image ($78$ of the $87$ images). This is the expected direction, since the spectral subtraction that produces the contrast image suppresses tissue that does not take
up iodine, so a finding that enhances weakly or not at all can appear unremarkable on the contrast image while remaining the established diagnosis for the breast. Because pathology is a breast-level property rather than an image-level one, these per-image disagreements reflect how a single underlying finding presents across acquisitions and views, not genuinely different diagnoses for the same tissue. We therefore harmonized the image labels to the most significant finding for each breast, ranking malignant over benign over normal, and applied the resulting breast-level label to both image types and both views. This ensures that every image of a given breast carries the same outcome label, so that the classifier is trained and evaluated against a consistent breast-level ground truth rather than against acquisition-dependent label noise.

\paragraph{Patient identifier and grouping variable.}
\label{sec:patient_grouping}
Each filename in CDD-CESM begins with a case identifier of the form \texttt{P100}, \texttt{P101}, and so on, corresponding to the patient from whom the images were acquired. We extract this identifier from the filename and use it as the patient-level grouping variable for all subsequent statistical operations. The identifier serves two distinct inferential roles: it is the partitioning unit for repeated nested cross-validation (Section~\ref{sec:cv}), where all images from a given
patient are assigned to the same outer fold so that no patient appears in both a training and a test partition, and it is the resampling unit for the patient-cluster bootstrap (Section~\ref{sec:bootstrap}), where the bootstrap draws patients with replacement and retains every image contributed by each drawn patient. The grouping variable is the mechanism through which the inference framework restricts both bias (via the cross-validation partition) and variance attribution (via the
bootstrap unit) to the patient level; image-level operations on either, given that each patient contributes a median of $8$ correlated images,
would conflate within-patient and between-patient variability.

The harmonization of the previous section operates within a single breast: it gives every image of a breast that breast's most significant label. It does not act across breasts, so a patient with disease in one
breast and a normal contralateral breast still carries two distinct breast-level labels; $184$ of the $308$ patients are of this mixed-label kind. Each binary task therefore needs a rule for reducing a patient's images to a single patient-level label, and that rule is the same for all four tasks. A task is named by two pathology classes; we keep only the images belonging to those two classes, drop the rest, and label a patient positive if any of her kept images belongs to the more severe class and negative otherwise. For normal-versus-malignant, for example, benign images are dropped and a patient is labeled $Y_s = 1$ if any of her remaining images is malignant and $Y_s = 0$ if they are all normal. One consequence of dropping the off-task class is worth stating: the $57$ patients whose every image is benign have no normal or malignant image, so they do not enter the normal-versus-malignant task at all and are never scored as negatives. Subject to this class restriction, no patient is excluded on account of holding mixed labels across her two breasts.

We use the any-positive rule rather than majority voting because it matches the clinical decision being modeled -- detection of the positive class in at least one breast, consistent with the worst-finding basis of BI-RADS assessment~\citep{ACRBI-RADS2013, zonderland_birads_2014} -- and because majority voting can assign a negative label to a patient with a genuine unilateral positive finding whenever her negative images outnumber her positive ones.

\section{Masked complex NDWT feature representation}
\label{sec:methods}

\subsection{Tissue mask and multiscale erosion}
\label{sec:tissue_mask}

The masked complex non-decimated wavelet feature bank is constructed in two stages: a tissue mask that defines the anatomical support over which all features are computed, and a complex non-decimated wavelet decomposition whose coefficients are summarized inside that support.

Let $A$ denote a mammogram, represented as an $m_1 \times m_2$ matrix of gray-level intensities. The black background surrounding the imaged breast is not part of the anatomy and must be excluded from every feature computation, since otherwise spurious zero values
inflate counts, deflate means, and bias higher moments of the coefficient distributions. We therefore construct a binary tissue mask
\begin{eqnarray}
M(i,k) =
\begin{cases}
1, & \text{if pixel } (i,k) \text{ belongs to breast tissue},\\
0, & \text{if pixel } (i,k) \text{ belongs to the outside background},
\end{cases}
\label{eq:tissue_mask}
\end{eqnarray}
obtained by estimating the background level from the image border, thresholding above this level, retaining the largest connected foreground component, filling interior holes, and applying mild morphological smoothing. The tissue support is
\begin{equation}
\Omega = \{(i,k) : M(i,k) = 1\}.
\label{eq:tissue_set}
\end{equation}
and every summary defined in the feature bank below is computed as an empirical average over $\Omega$ rather than over the full image grid. This is the statistical-domain restriction that makes the feature bank scale-aware: it ensures that masked log-energies, masked quantiles, and masked phase summaries reflect the distribution of wavelet coefficients on breast tissue alone, not on a mixture of tissue and background that depends on the per-image size of $\Omega^c$.

A single image-domain mask, however, is not sufficient on its own. The wavelet detail coefficients at level $j$ are computed from a filter whose impulse response spans a finite spatial extent, so a coefficient at $(i,k)$ depends on pixels of $A$ within a neighborhood whose radius grows with $j$. Coefficients near the boundary of $\Omega$ therefore mix contributions from breast tissue and from the background, and this cone-of-influence contamination
becomes worse at coarser scales. To control it, we erode $M$ at each scale to produce a level-dependent mask $M_e^{(j)}$ whose retained pixels are guaranteed to be far enough from the boundary that the level-$j$ wavelet coefficients depend only on tissue values. Erosion is performed with a structuring element matched to
the support of the wavelet filter at level $j$; in our experiments we use $L = 7$ decomposition levels. The multiscale erosion is a deliberate design choice rather than a preprocessing convenience: without it, every coarse-scale feature would carry boundary-driven bias whose magnitude varies systematically with the size and shape
of the breast in each mammogram, confounding the scale-decay signatures that several of the feature families described below are constructed to capture.

\subsection{Complex non-decimated wavelet transform}
\label{sec:ndwt}

The masked image is then decomposed using a two-dimensional complex non-decimated wavelet transform (NDWT). In the experiments reported below we use a complex-valued Daubechies filter with coefficients given in Supplement~S2 and in the released
code. Circular boundary handling is used in the non-decimated convolution so that all
coefficient images remain on the original image grid. At each level $j = 1, \ldots, L$, the transform produces three directional detail subbands and one approximation subband:
\begin{eqnarray}
D_j = HH_j, \qquad H_j = HL_j, \qquad V_j = LH_j,
\label{eq:subbands}
\end{eqnarray}
corresponding to diagonal, horizontal, and vertical detail content, together with the final smooth image $S_L$. Two properties of this transform matter statistically for the feature bank that follows.

First, the transform is non-decimated: the output coefficient image at each level retains the full $m_1 \times m_2$ resolution of the input, rather than being downsampled by a factor of two per axis at each successive level as in the classical (decimated) wavelet transform. Every coefficient image $C_{j,s}$ with $s \in \{D, H, V\}$ therefore has
size $m_1 \times m_2$, identical to that of $A$. The level-eroded tissue mask $M_e^{(j)}$ therefore applies pointwise to its coefficient image at every scale, without resampling or interpolation: the mask grid and the coefficient grid coincide at every level, so only the erosion radius changes with $j$, not the sampling. This pointwise alignment is what makes the masking operation well-defined as a statistical restriction: a
masked decimated wavelet transform requires upsampling the mask to match each scale's coefficient grid, which introduces interpolation artifacts and a per-scale ambiguity in which pixels belong to the support. The non-decimated representation avoids that ambiguity entirely.

Second, because the filter is complex-valued, each coefficient is itself a complex number. The magnitude $|C_{j,s}|$ carries energy and shape information and is the analogue of the real-valued wavelet coefficient used in classical wavelet feature banks; the phase $\arg C_{j,s}$ carries information about local oriented structure --- specifically, the alignment and coherence of edges, ridges, and other directional content
within each subband at each scale. Real-valued wavelet transforms discard the phase channel altogether, and so cannot represent phase-coherence statistics; the complex non-decimated transform makes phase available as a separate channel of the feature bank, on equal footing with magnitude.

For any coefficient image $C_{j,s}$, the masked working vector is
restricted to the level-eroded tissue support
\begin{equation}
\Omega_j = \{(i,k) : M_e^{(j)}(i,k) = 1\},
\label{eq:level_support}
\end{equation}
the set of pixels retained by the level-$j$ eroded mask of
Section~\ref{sec:tissue_mask}, giving
\begin{eqnarray}
x_{j,s} = \{C_{j,s}(i,k) : (i,k) \in \Omega_j\},
\label{eq:masked_vector}
\end{eqnarray}
with magnitudes $r_{j,s} = |x_{j,s}|$ and phases
$\phi_{j,s} = \arg(x_{j,s})$. All energy descriptors use $|x|^2$, not
$x^2$, since the coefficients are complex. The masked log-energy at
level $j$ in subband $s$ is
\begin{equation}
E_{j,s}
= \frac{1}{|\Omega_j|}
\sum_{(i,k) \in \Omega_j} |C_{j,s}(i,k)|^2,
\qquad
L_{j,s} = \log_2 E_{j,s},
\label{eq:masked_log_energy}
\end{equation}
and analogous masked averages define all higher-order moments,
quantiles, and phase summaries used downstream. The key point is that every sum is taken over the level-appropriate eroded tissue support $\Omega_j$, never over the full image grid.

\begin{figure}[!htbp]
\centering
\begin{tikzpicture}[
    line width=0.9pt,
    >=Latex,
    font=\large
]

\def\imgw{1.45cm}
\def\imgh{2.25cm}
\def\rowsep{3.05}   
\def\colsep{1.75}   

\tikzset{imgbox/.style={draw, inner sep=0pt}}

\node[imgbox] (Abox) at (-5.6,1.30)
    {\includegraphics[width=\imgw,height=\imgh,keepaspectratio]{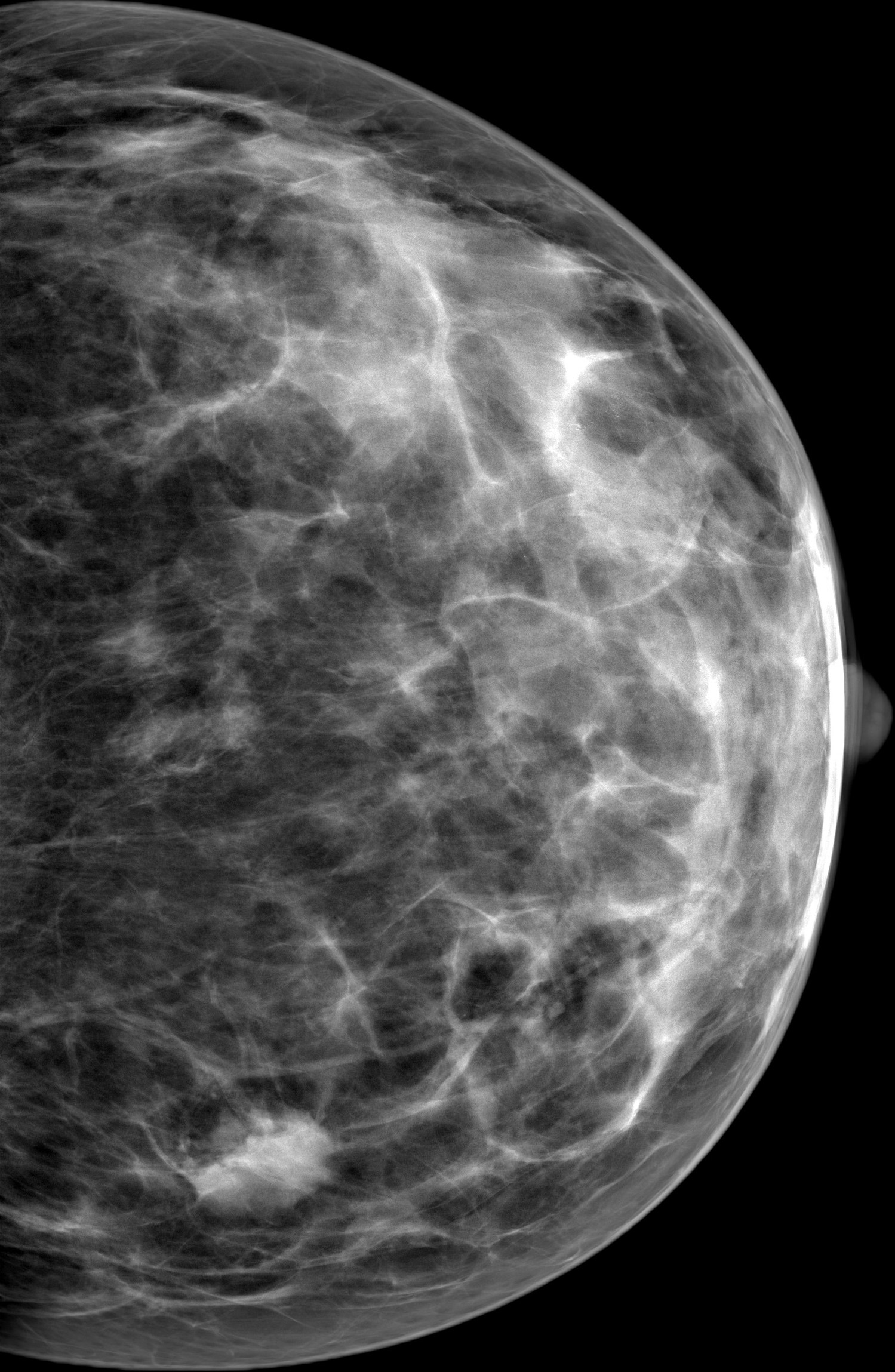}};
\node[below=2pt of Abox] {$A$};

\node[imgbox] (Mbox) at (-5.6,4.55)
    {\includegraphics[width=\imgw,height=\imgh,keepaspectratio]{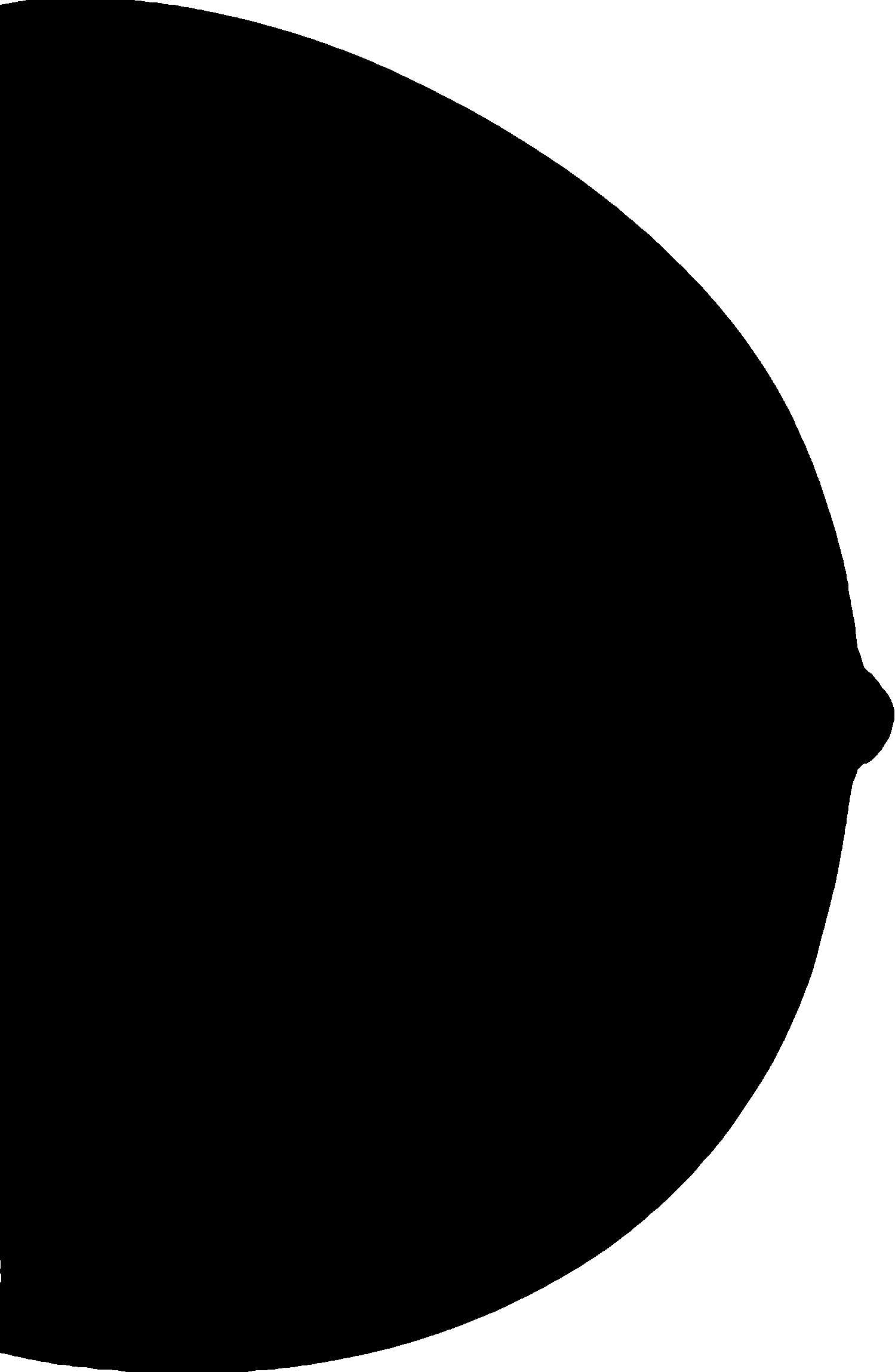}};
\node[above=2pt of Mbox] {$M$};

\node[imgbox] (MEbox) at (-3.85,4.55)
    {\includegraphics[width=\imgw,height=\imgh,keepaspectratio]{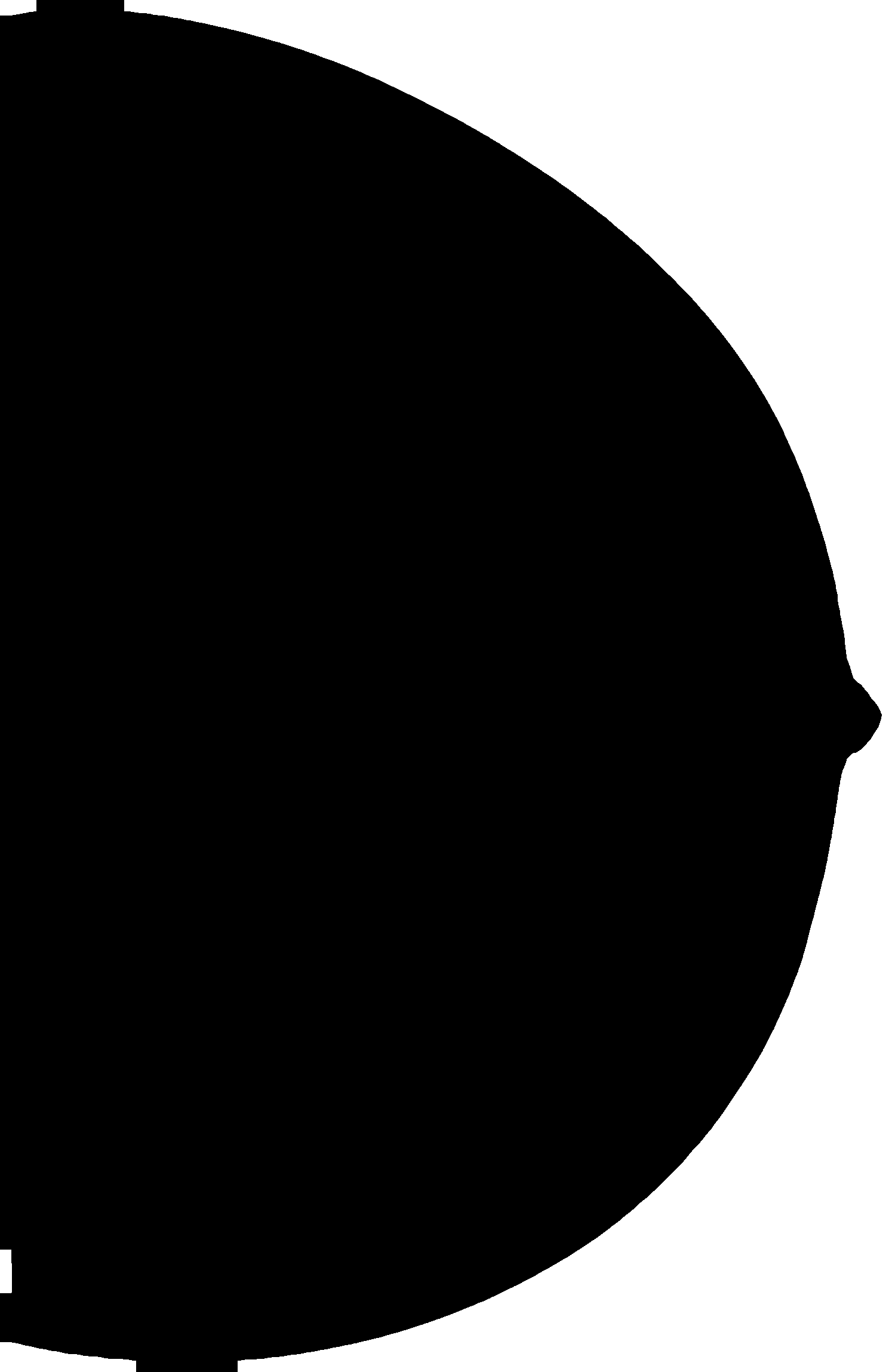}};
\node[above=2pt of MEbox] {$M_e$};

\draw[->,line width=1.1pt] (-3.0,2.90) -- (-0.55,2.90);
\node[above=1pt,align=center,font=\bfseries] at (-1.78,2.90) {masked\\ND2D WT};

\newcommand{\ndtile}[4]{%
    \node[imgbox] (#1) at (#2,#3)
        {\includegraphics[width=\imgw,height=\imgh,keepaspectratio]{#4}};%
}

\node[imgbox] (S3) at (0.55,5.30) {\includegraphics[width=\imgw,height=\imgh,keepaspectratio]{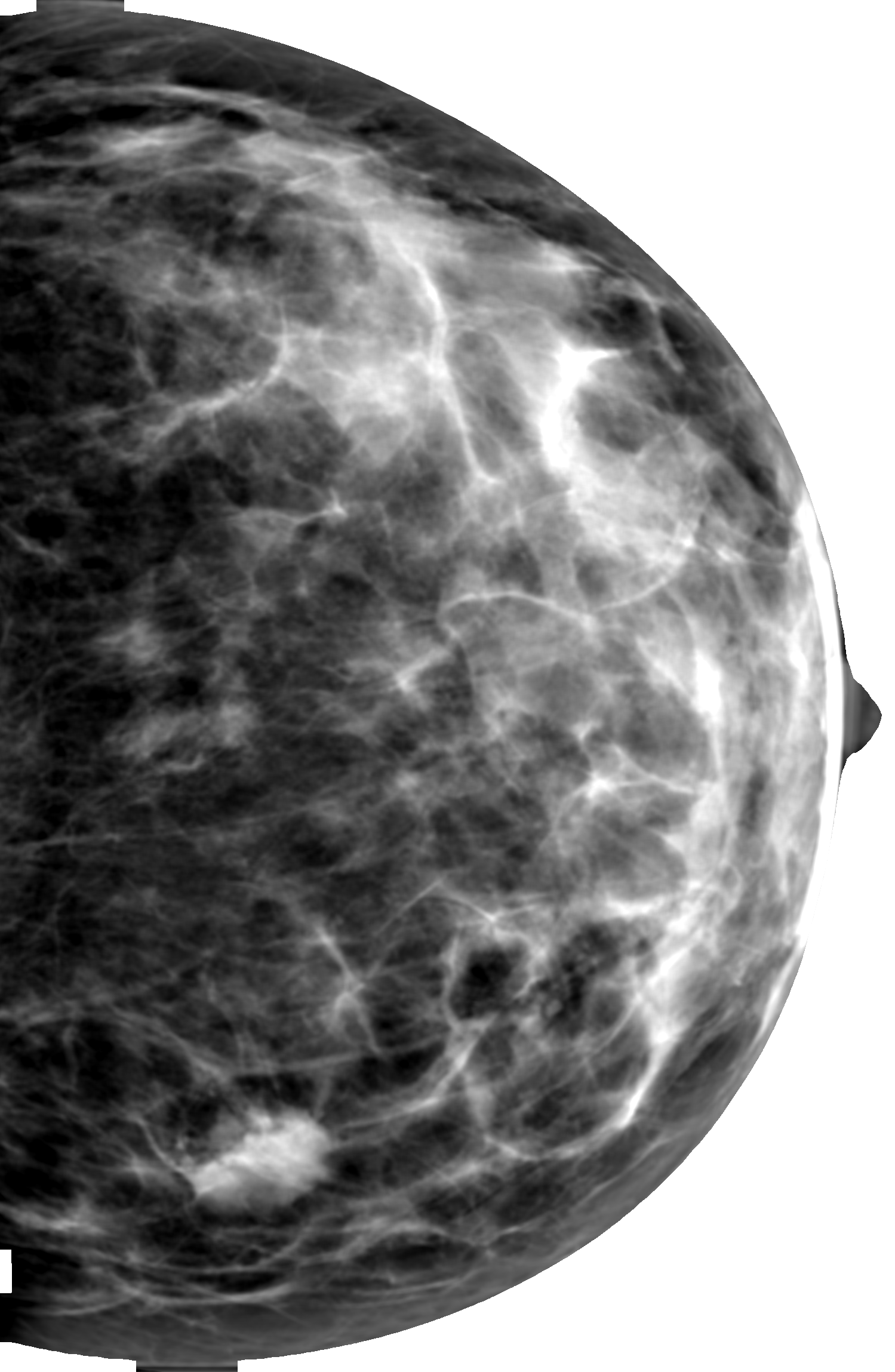}};  \node[above=2pt of S3]{$S_3$};
\node[imgbox] (H3) at (2.30,5.30) {\includegraphics[width=\imgw,height=\imgh,keepaspectratio]{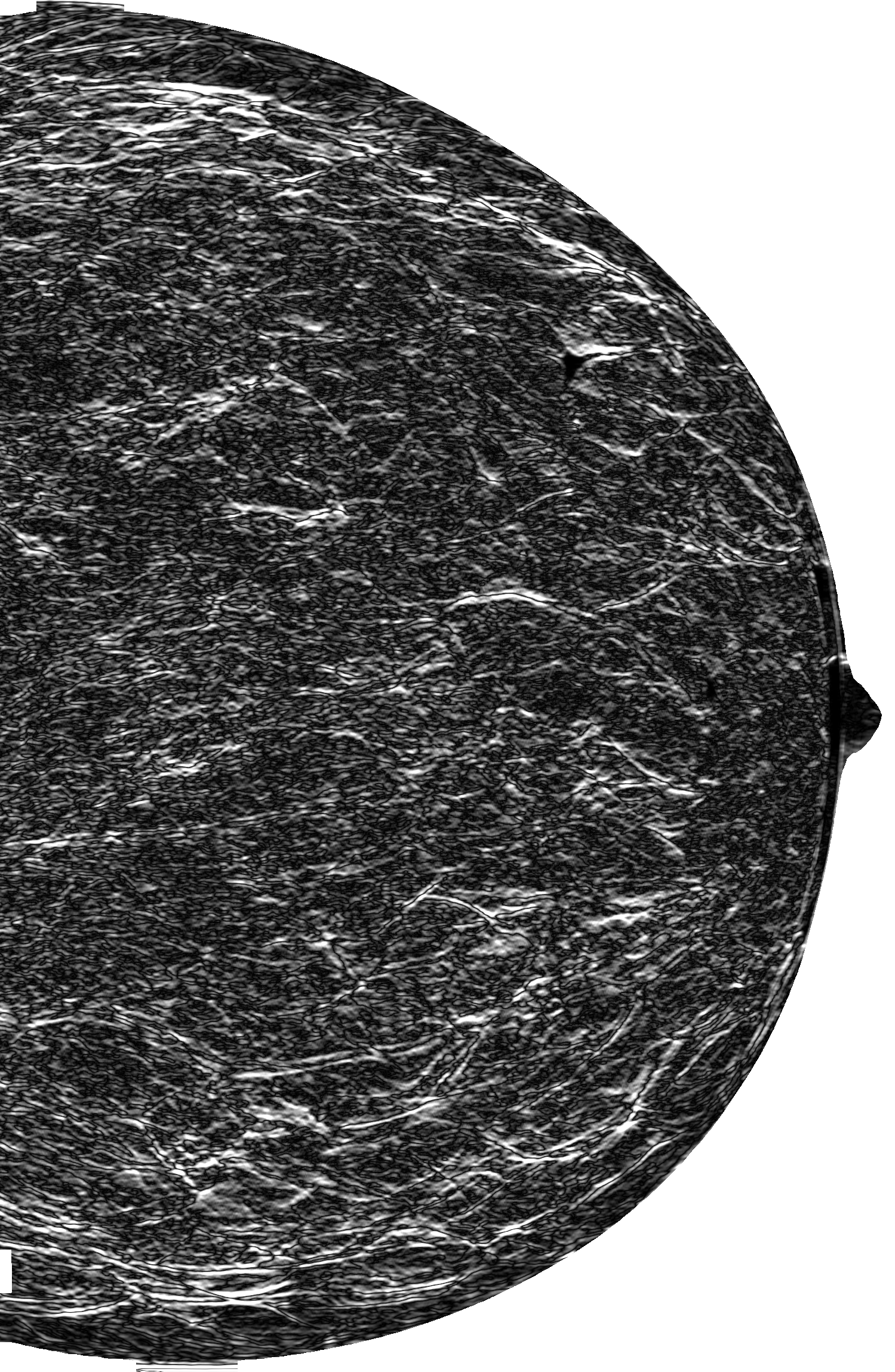}};  \node[above=2pt of H3]{$H_3$};

\node[imgbox] (V3) at (0.55,2.25) {\includegraphics[width=\imgw,height=\imgh,keepaspectratio]{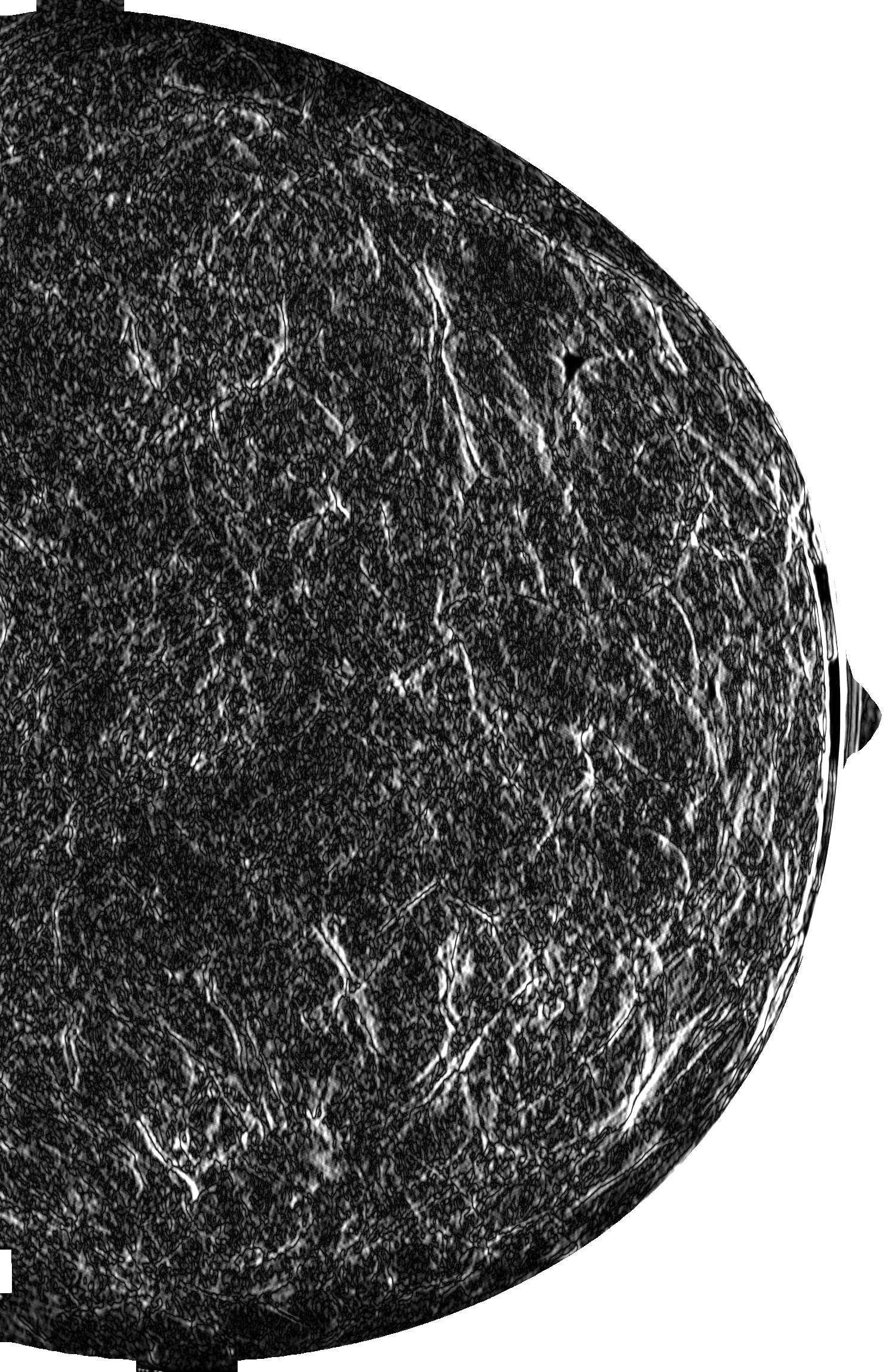}};  \node[above=2pt of V3]{$V_3$};
\node[imgbox] (D3) at (2.30,2.25) {\includegraphics[width=\imgw,height=\imgh,keepaspectratio]{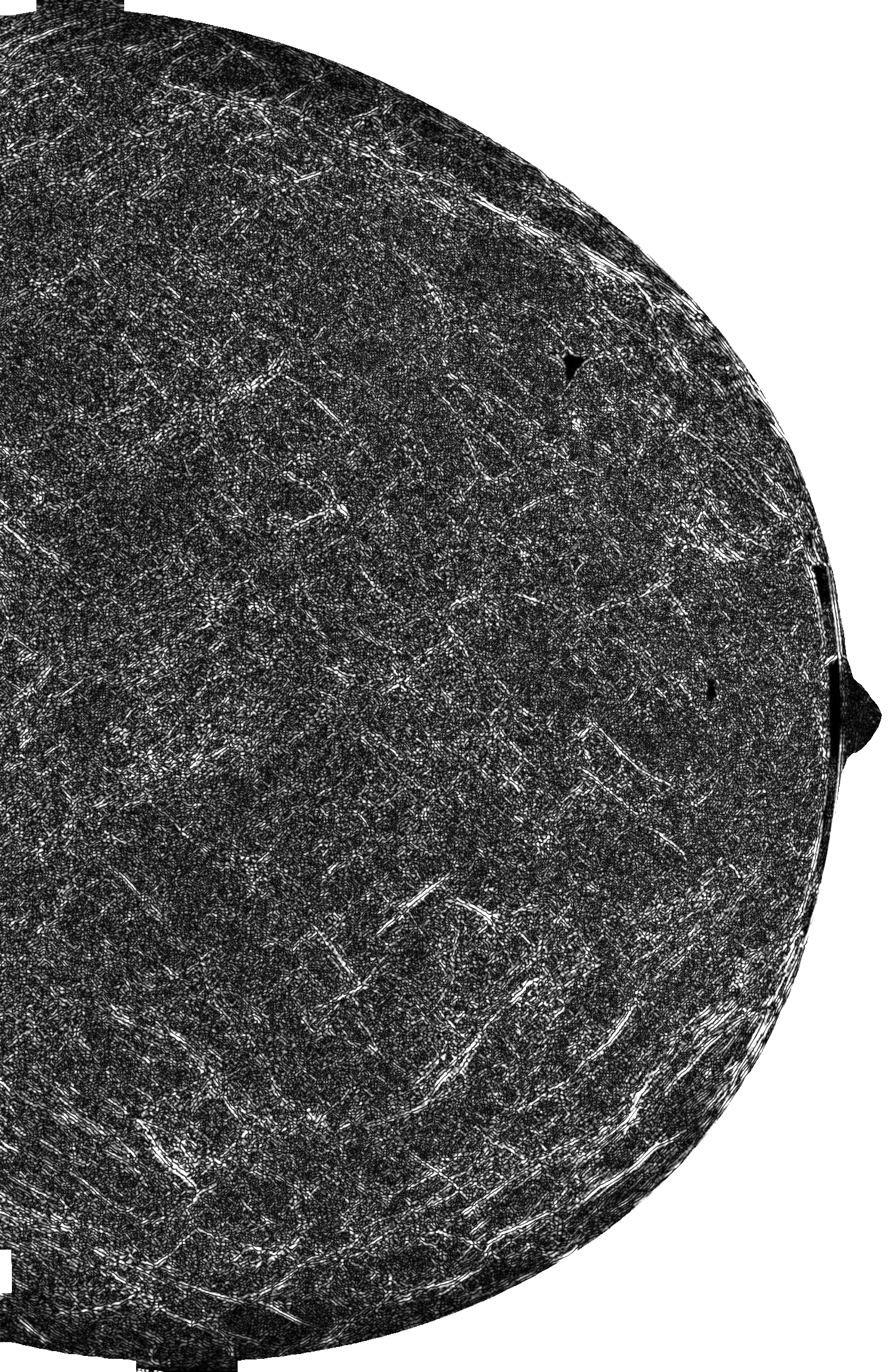}};  \node[above=2pt of D3]{$D_3$};
\node[imgbox] (H2) at (4.05,2.25) {\includegraphics[width=\imgw,height=\imgh,keepaspectratio]{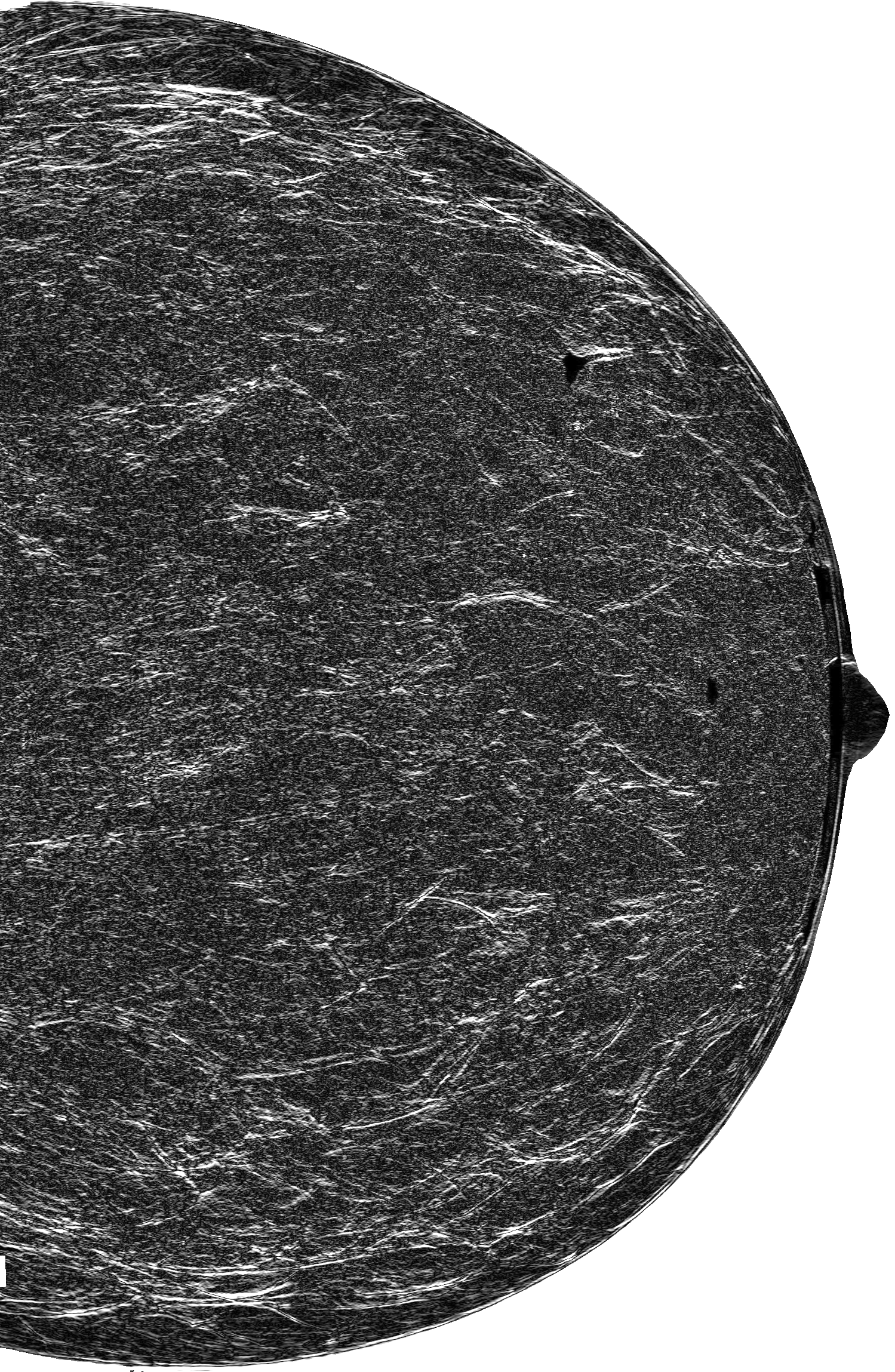}};  \node[above=2pt of H2]{$H_2$};

\node[imgbox] (V2) at (2.30,-0.80) {\includegraphics[width=\imgw,height=\imgh,keepaspectratio]{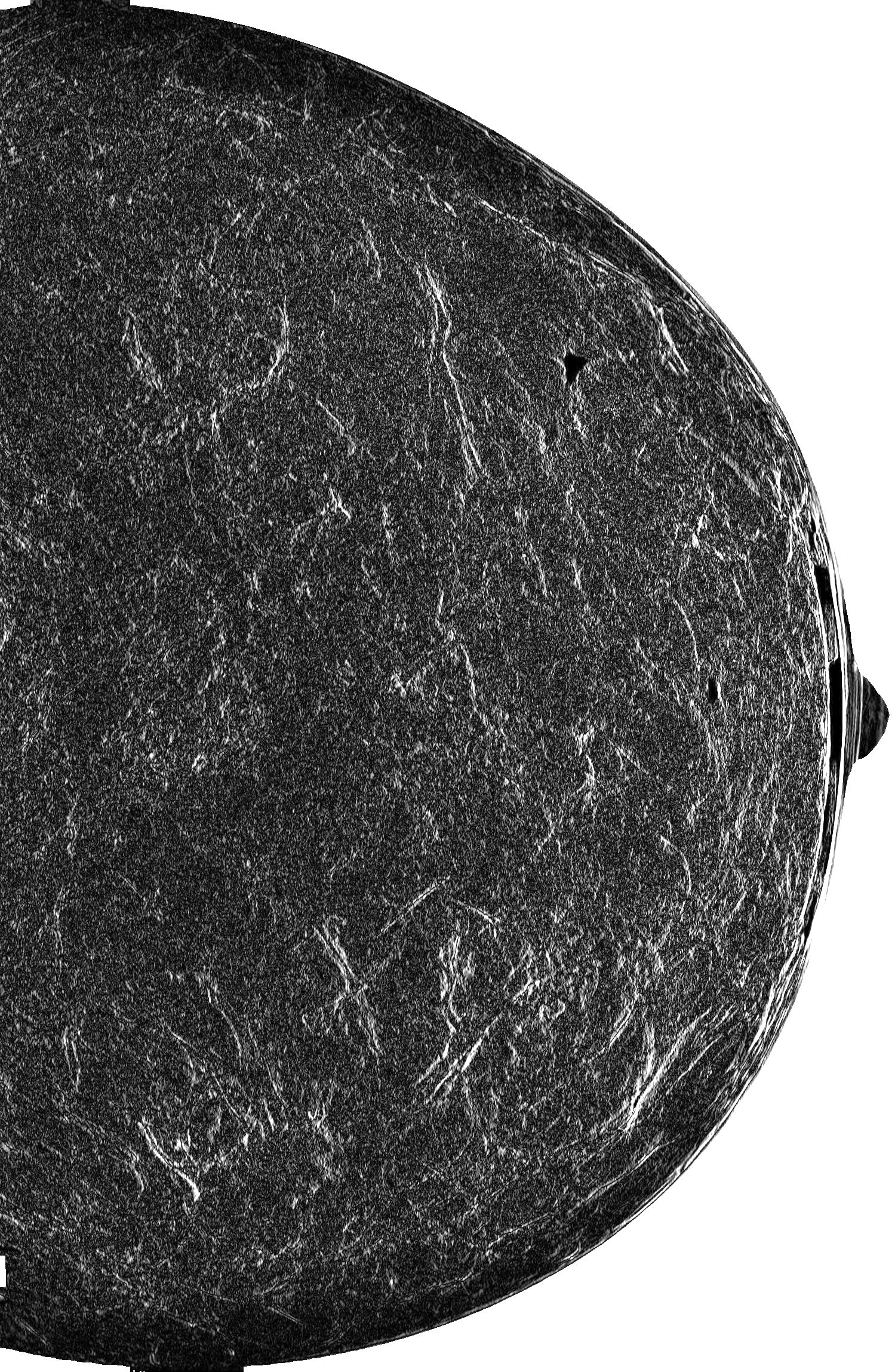}}; \node[above=2pt of V2]{$V_2$};
\node[imgbox] (D2) at (4.05,-0.80) {\includegraphics[width=\imgw,height=\imgh,keepaspectratio]{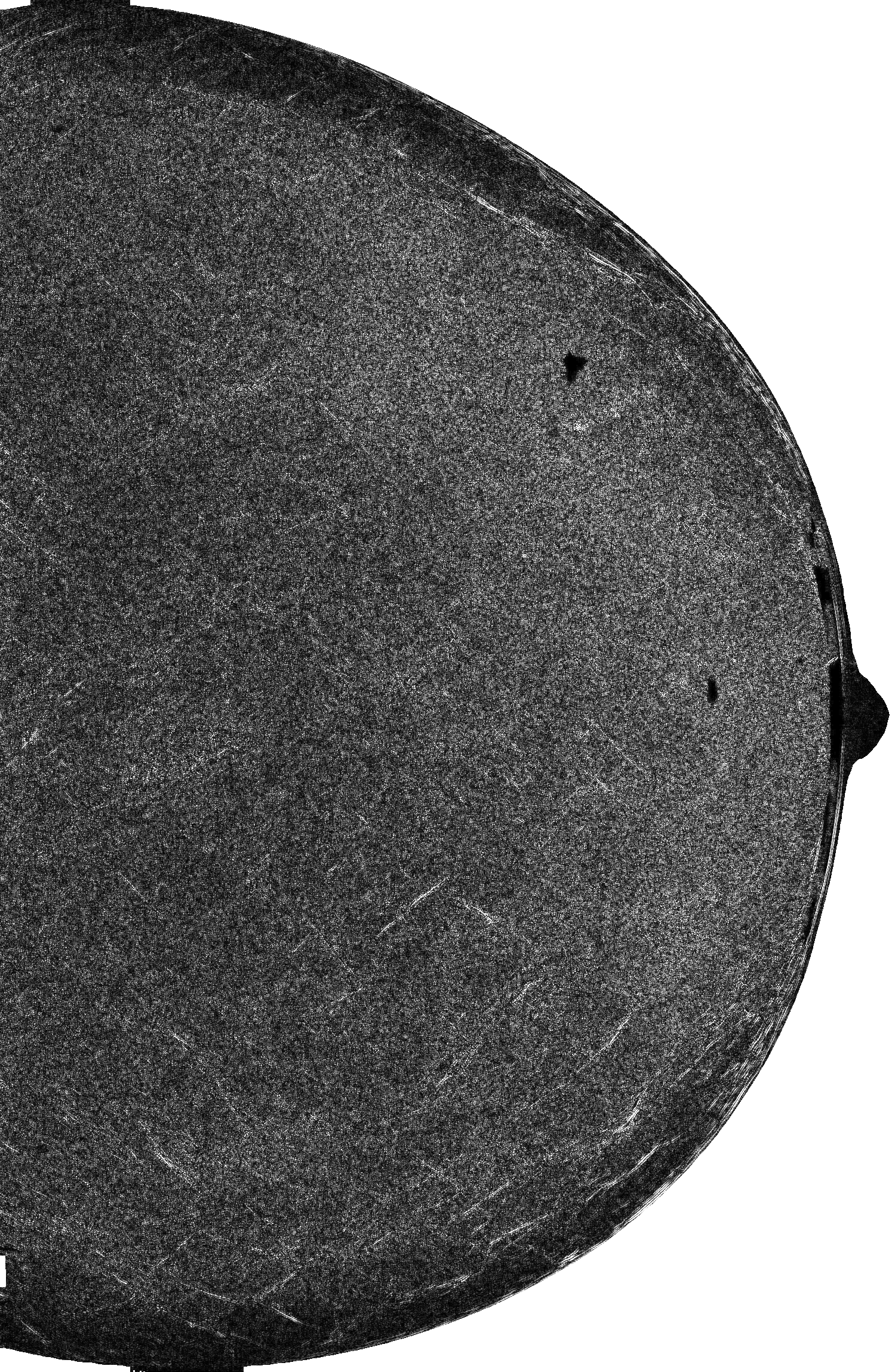}}; \node[above=2pt of D2]{$D_2$};
\node[imgbox] (H1) at (5.80,-0.80) {\includegraphics[width=\imgw,height=\imgh,keepaspectratio]{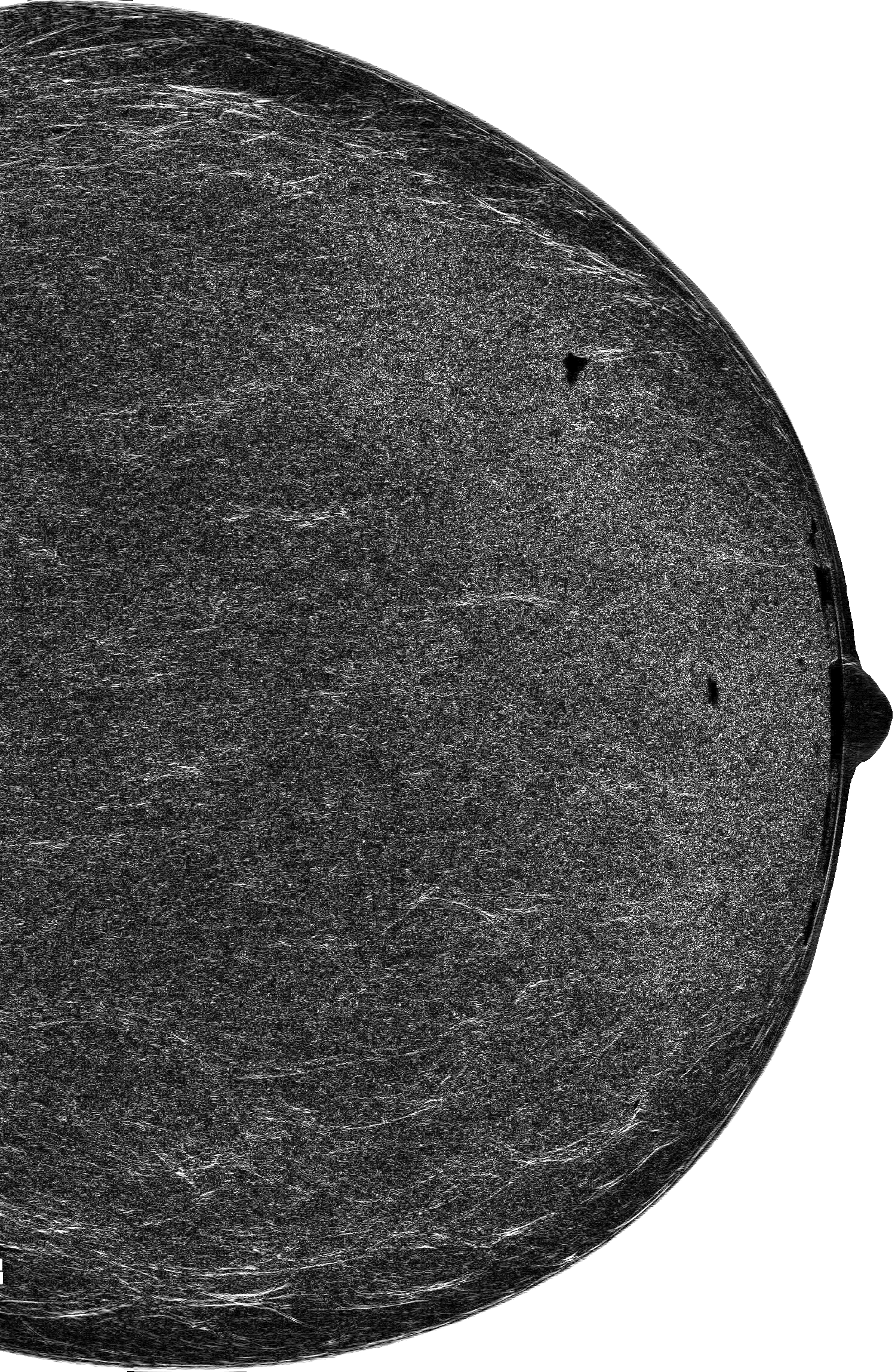}}; \node[above=2pt of H1]{$H_1$};

\node[imgbox] (V1) at (4.05,-3.85) {\includegraphics[width=\imgw,height=\imgh,keepaspectratio]{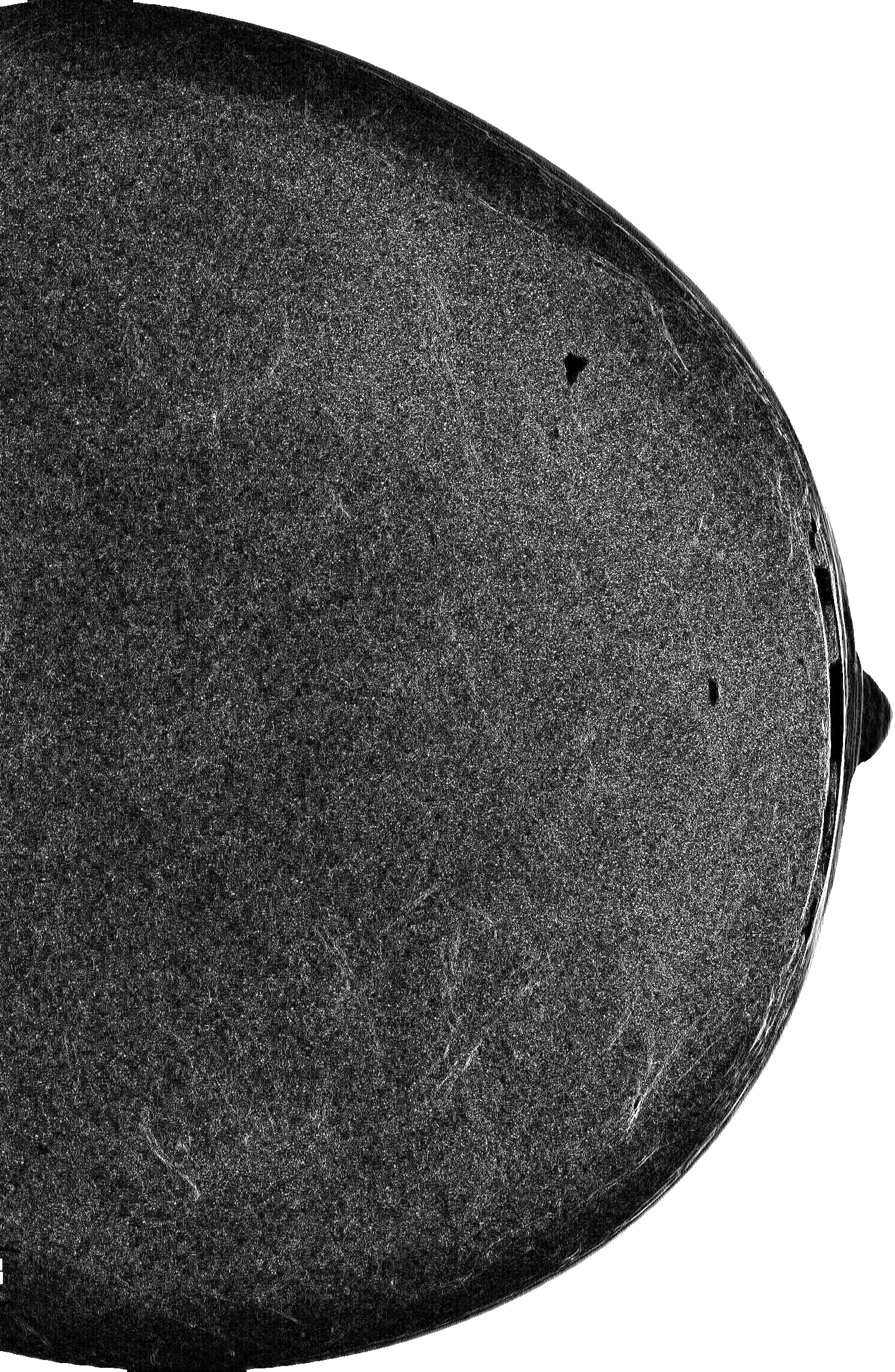}}; \node[above=2pt of V1]{$V_1$};
\node[imgbox] (D1) at (5.80,-3.85) {\includegraphics[width=\imgw,height=\imgh,keepaspectratio]{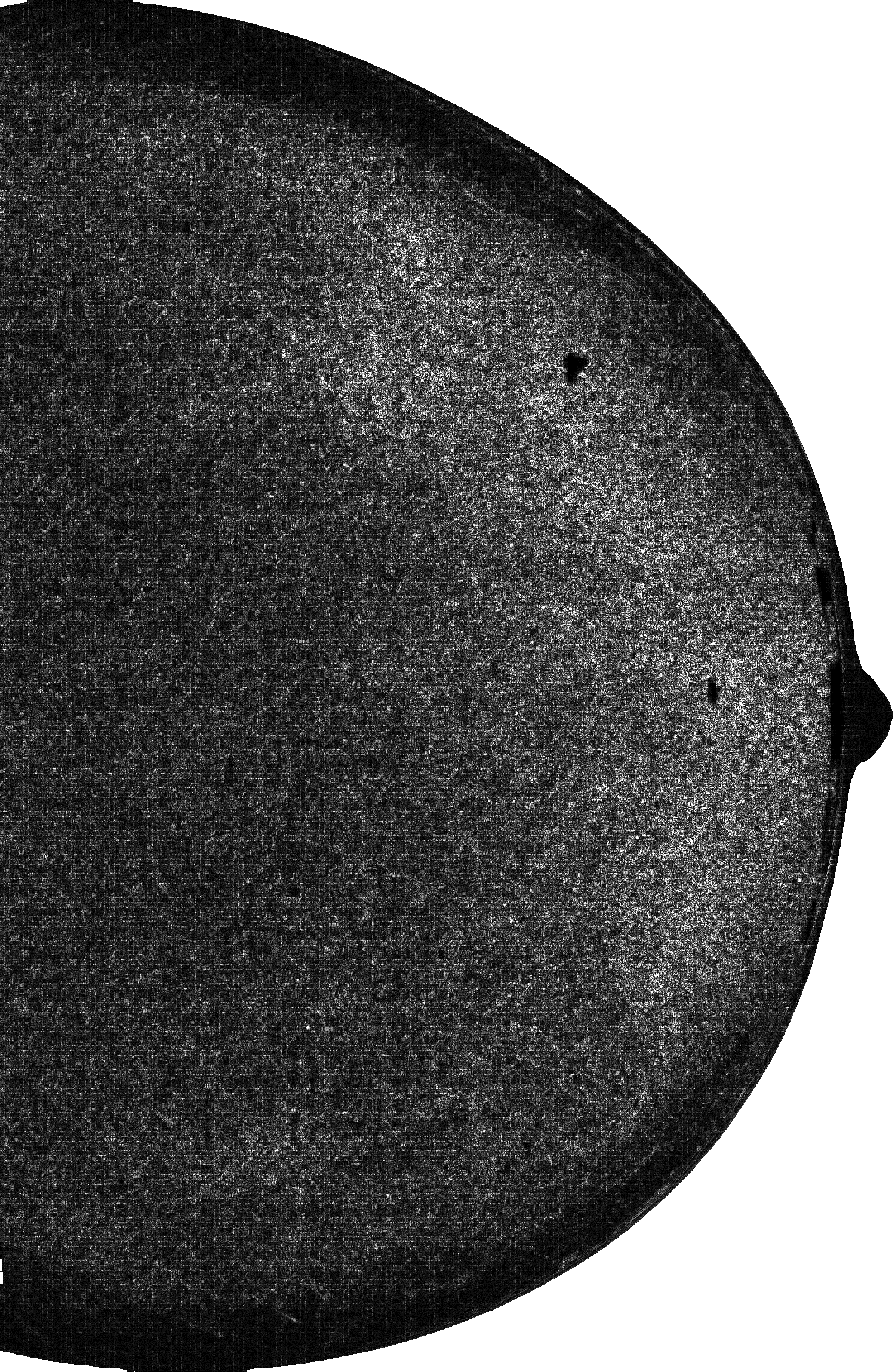}}; \node[above=2pt of D1]{$D_1$};

\node[align=center,font=\normalsize] at (3.0,-5.55)
{All coefficient images have size $m_1\times m_2$, identical to the input.};

\end{tikzpicture}

\caption{Masked non-decimated two-dimensional wavelet decomposition of a
mammogram. On the left are the original mammogram $A$, the tissue mask
$M$, and a representative eroded mask $M_e$. On the right are the masked
smooth and detail coefficient images. Because the transform is
non-decimated, every coefficient image has the same size as $A$, so the
mask applies pointwise at every scale without resampling; the erosion
itself is level-dependent (the mask $M_e^{(j)}$ of
Section~\ref{sec:tissue_mask}), and the $M_e$ panel at left shows one
eroded mask for reference.}
\label{fig:ndwt_mammo_schematic}
\end{figure}

Figure~\ref{fig:ndwt_mammo_schematic} illustrates the construction described in Sections~\ref{sec:tissue_mask} and~\ref{sec:ndwt} on a representative mammogram: the original image $A$, the tissue mask $M$, and a representative eroded mask $M_e$ are shown on the left, and the
masked smooth and detail coefficient images $S_3$, $H_j$, $V_j$, $D_j$ for $j = 1, 2, 3$ are shown on the right. Because the transform is non-decimated, every coefficient image has the same spatial size as $A$, so the mask applies pointwise at every scale. An analogous display on the Lena test image is provided in Supplement~S1 to make the multi-scale subband organization more transparent on a familiar non-clinical example.

\subsection{Feature families}
\label{sec:features}

From the masked complex NDWT representation we extract $203$ numeric features per image, organized into seven families summarized in Table~\ref{tab:feature_families}. The families together span the kinds of image content that a multiscale representation can summarize: the marginal distribution of energy across scales (log-energy), the shape and tail behavior of the coefficient magnitude distribution at each scale (magnitude distribution), the relative balance among the three directional subbands (anisotropy and dominance), the coherence of oriented structure that magnitude alone cannot see (phase and coherence), the cross-scale persistence of strong responses (persistence and cross-scale), the rate at which energy decays with scale (energy spectrum slopes), and the level-to-level energy differences that this decay produces (adjacent-scale log-energy differences). A complete feature codebook with exact mathematical definitions and the naming convention used in the released code is provided in Supplement~S2.

\begin{table}[!htbp]
\centering
\caption{Feature families extracted from the masked complex
non-decimated wavelet representation of each mammogram. All summaries
are computed over the level-eroded tissue support $\Omega_j$ of
Section~\ref{sec:ndwt}.  See Supplement~S2 for the full codebook.}
\label{tab:feature_families}
\small
\begin{tabular}{p{0.27\textwidth} p{0.66\textwidth}}
\toprule
Family & Descriptors \\
\midrule
Wavelet log-energy &
Per-direction, per-level $\log_2$ energy of the masked subband, with
energy-concentration coefficients and energy-entropy summaries. \\
Magnitude distribution &
Coefficient of variation, kurtosis, upper-tail quantiles, and
tail-fraction summaries of $|C_{j,s}|$. \\
Anisotropy / dominance &
Per-level fractional energy in each directional channel, dominance
entropy, and log-ratios between horizontal, vertical, and diagonal
channels. \\
Phase / coherence &
Per-level mean resultant lengths of $\arg C_{j,s}$ within each
directional channel and inter-channel phase coherence summaries
(e.g.\ circular correlation between $\arg H_j$ and $\arg V_j$).
Available only because the transform is complex-valued. \\
Persistence / cross-scale &
Adjacent-scale correlations between levels, maximum over levels,
and level-of-maximum summaries. \\
Energy spectrum slopes &
Linear slopes and $R^2$ values from regressing $\log_2 E_{j,s}$ on
level $j$ over selected ranges, capturing how energy decays with
scale. \\
Adjacent-scale log-energy differences &
Differences of the form $L_{j+1,s} - L_{j,s}$. \\
\bottomrule
\end{tabular}
\end{table}

Several of these families have direct radiological correspondences that motivate the feature-selection patterns reported in Section~\ref{sec:res-features}. Phase coherence indexes the consistency of local oriented structure and is therefore sensitive to architectural distortion and spiculation, which are the dominant low-energy mammographic cues for malignancy in the BI-RADS
lexicon~\citep{ACRBI-RADS2013}. Magnitude-distribution descriptors index how concentrated wavelet energy is in a small fraction of the tissue support; on the recombined contrast image, where the spectral subtraction has already suppressed background tissue, this concentration corresponds to localized iodine uptake from tumor
neovasculature. Energy-spectrum slopes and adjacent-scale
differences describe the rate at which tissue texture coarsens with scale, summarizing the scale-invariance properties of breast parenchyma that are clinically reported as breast density. Anisotropy and persistence descriptors capture, respectively, the directional organization of tissue and the multi-scale support of
localized structures such as masses, distinguishing them from isotropic background parenchyma and from scale-specific noise. The correspondences motivate the feature-selection analysis in Section~\ref{sec:res-features}.

\section{Modeling and inference framework}
\label{sec:model}

With each mammogram now represented as a fixed-length vector of
masked multiscale descriptors, we turn to the predictive model that maps these features to a probability of malignancy, the
cross-validation framework used to evaluate it, the rules used to aggregate image-level predictions into patient-level predictions, and the bootstrap procedure used for inference on the resulting patient-level AUCs.

\subsection{Elastic-net classifier and nonlinear comparators}
\label{sec:elasticnet}

For each binary task we model the class probability with a
regularized logistic regression. Let $y_i \in \{0,1\}$ denote the image-level outcome and $x_i \in \mathbb{R}^p$ the corresponding feature vector extracted from the masked complex NDWT representation. The classifier estimates the conditional class probability
\begin{equation}
  \Pr(Y_i = 1 \mid x_i)
  \;=\;
  \frac{1}{1 + \exp\bigl\{-(\beta_0 + x_i^\top \beta)\bigr\}},
\end{equation}
with the coefficient vector chosen by penalized maximum likelihood:
\begin{equation}
  \hat{\beta} \;=\; \arg\min_{\beta_0, \beta}\;\left\{
    -\frac{1}{n}\sum_{i=1}^n \left[ y_i (\beta_0 + x_i^\top \beta)
      - \log\bigl(1 + e^{\beta_0 + x_i^\top \beta}\bigr) \right]
    + \lambda \left[ \alpha \|\beta\|_1
                   + (1-\alpha) \tfrac{1}{2}\|\beta\|_2^2 \right]
  \right\}.
  \label{eq:enet}
\end{equation}
The penalty in equation~\eqref{eq:enet} is the elastic-net penalty of~\citet{zou_hastie_elasticnet_2005}, a convex combination of the $\ell_1$ lasso~\citep{tibshirani_lasso_1996} and the $\ell_2$ ridge penalty~\citep{hoerl_kennard_ridge_1970}. The mixing parameter
$\alpha \in [0,1]$ interpolates between pure ridge ($\alpha = 0$), which shrinks coefficients but retains all predictors, and pure lasso ($\alpha = 1$), which produces a sparse model by zeroing some coefficients exactly. The choice of an elastic-net penalty rather than a pure lasso is dictated by the structure of the feature bank.
The $203$ masked complex NDWT descriptors are heavily correlated
within and across scale and direction: log-energies at adjacent
levels within the same directional subband move together, and
persistence, energy-spectrum-slope, and adjacent-scale-difference features are algebraic combinations of the same underlying $E_{j,s}$. In this regime the pure lasso is known to behave erratically,
arbitrarily selecting one variable from a correlated group and discarding the others; the ridge component of the elastic-net penalty stabilizes selection by encouraging strongly correlated predictors to enter or leave the model together, the so-called grouping effect of
\citet{zou_hastie_elasticnet_2005}. We sweep $\alpha$ over a
small grid $\{0.3, 0.5, 0.7\}$ rather than fixing a single value, since the optimal lasso/ridge balance is task-dependent and is more usefully tuned than guessed; selection across the grid is done by inner cross-validation, as described in Section~\ref{sec:cv}.

The regularization strength $\lambda$ is selected for each $\alpha$ by inner $K_\text{inner} = 5$-fold cross-validation on each outer training set, using the one-standard-error rule of
\citet{breiman_olshen_stone_1984}: among the candidate values of
$\lambda$ whose cross-validated deviance is within one standard
error of the minimum, we choose the most regularized. The 1SE rule is preferred over the minimum-deviance rule because the minimum is typically a flat, noisy region of the cross-validation curve, so choosing the smallest $\lambda$ in that flat region produces a sparser and more stable selection without sacrificing predictive performance~\citep{hastie_tibshirani_friedman_2009}.

To test whether the modality and fusion findings are specific to the linear elastic-net classifier rather than to the masked complex NDWT feature bank itself, we additionally evaluate two nonlinear classifiers under the identical patient-grouped pipeline: a support vector machine with a radial basis function kernel (\emph{svm\_radial}) and a gradient-boosted tree ensemble (\emph{gboost}). Both are fitted on the same masked complex NDWT feature vectors, partitioned by the same
patient-grouped outer folds, with the same in-fold median imputation and standardization described in Section~\ref{sec:cv}, and their hyperparameters are tuned by the same inner five-fold cross-validation used for the elastic-net path. Image-level out-of-fold probabilities from each classifier are aggregated to the patient level by the four
fusion rules of Section~\ref{sec:fusion} and summarized by the
patient-cluster bootstrap of Section~\ref{sec:bootstrap}, so that the three classifiers differ only in the function class mapping features to probabilities and not in any element of the evaluation framework.

\subsection{Patient-grouped nested cross-validation}
\label{sec:cv}

The classifier of Section~\ref{sec:elasticnet} is evaluated under repeated patient-grouped nested cross-validation. The outer loop uses $K = 5$ folds and $R = 10$ independent repeats, and the inner loop uses $K_{\text{inner}} = 5$ folds. Three properties of this scheme are statistically substantive and worth stating explicitly.

\paragraph{Patient-level partitioning.} The grouping variable introduced in Section~\ref{sec:patient_grouping} is used to assign every image contributed by a single patient to the same outer fold. In each repeat we partition the $308$ patients into five approximately equal
groups, stratified on the any-positive patient label $Y_s$ of Section~\ref{sec:patient_grouping}, so that no patient appears in both a training and a test partition and the class balance of the
task is preserved across folds. Image-level cross-validation, in which images from the same patient can appear in both training and test partitions, measures the classifier's ability to predict on new images of patients it has already seen. Patient-grouped cross-validation measures the harder and more relevant quantity: prediction on new patients. The gap between the two can be
substantial~\citep{yagis_data_leakage_2021, rouzrokh_mitigating_2022,
varoquaux_machine_2022}; the patient-grouped partition is what
closes it.

\paragraph{In-fold preprocessing.} Two preprocessing steps are applied within each outer fold before the elastic-net path is fitted. Features with missing values are imputed using the per-feature median computed on the training portion of that outer fold only; all features are then centered and scaled to unit variance using the same training-fold-only statistics. Computing the imputation and
standardization statistics inside each outer fold, rather than once on the full dataset, is essential: any preprocessing that uses information from the held-out test images leaks information across
the split and biases performance estimates
upward~\citep{kapoor_narayanan_leakage_2023,
ambroise_mclachlan_2002}. This is a less visible form of leakage than the partition itself, but it has the same effect, and the patient-grouped partition does not by itself prevent it.

\paragraph{Inner-loop model selection.} Within each outer training set, the inner $K_{\text{inner}} = 5$-fold cross-validation is used to select $\lambda$ for each candidate $\alpha$ on the elastic-net grid, again using the one-standard-error rule of
Section~\ref{sec:elasticnet}. Tying model selection to an inner cross-validation, rather than to the outer test folds, prevents the selection of $\lambda$ from being informed by performance on
the outer test images and is a standard requirement of nested cross-validation~\citep{varoquaux_machine_2022}.

The outer fold yields out-of-fold (OOF) predicted probabilities for each test image. The OOF probabilities are averaged across the $R$
repeats to produce a single per-image prediction, and per-fold image-level AUCs are recorded across the $RK = 50$ outer-fold fits to characterize variability.

\subsection{Patient-level prediction and fusion}
\label{sec:fusion}

The classifier of Section~\ref{sec:elasticnet} produces a predicted probability $p_i \in [0,1]$ for each individual image. Clinical decisions, however, are made at the patient level: a screening mammogram is read as a whole study, and a patient is
either recalled for further work-up or returned to routine screening on the basis of the combined evidence from all of her acquired
images~\citep{ACRBI-RADS2013}. The classifier's image-level probabilities therefore have to be
combined into a single per-patient probability before patient-level performance can be assessed. This combination step, sometimes called multi-view or multi-image fusion, is methodologically consequential because different aggregation rules correspond to different implicit assumptions about how the per-image evidence should be combined, and they can yield substantially different patient-level performance from the same underlying image-level
predictions~\citep{carneiro_unregistered_2017, wu_deep_2020}.

Let $\mathcal{I}(s)$ denote the set of out-of-fold images for patient $s$, with associated probabilities $\{p_i : i \in \mathcal{I}(s)\}$. We compare four fusion rules spanning the spectrum from full averaging to extreme order-statistic selection.
The first, denoted \textit{raw mean}, is the arithmetic mean across all of the patient's images,
\begin{equation}
  \bar{p}_s^{\,\text{raw}}
  \;=\;
  \frac{1}{|\mathcal{I}(s)|}
  \sum_{i \in \mathcal{I}(s)} p_i.
  \label{eq:raw_mean}
\end{equation}
This treats the images as exchangeable noisy estimates of a single per-patient probability, and is the implicit fusion rule when image-level out-of-fold predictions are simply pooled by averaging.
The second rule, \textit{mean views}, averages within each of the two standard mammographic projections (CC and MLO) and then averages the two within-view means,
\begin{equation}
  \bar{p}_s^{\,\text{mv}}
  \;=\;
  \tfrac{1}{2}\bigl(\bar{p}_s^{\,\text{CC}} + \bar{p}_s^{\,\text{MLO}}\bigr).
  \label{eq:mean_views}
\end{equation}
This rule weights the two projections equally regardless of how many images of each are available. The third rule, \textit{max views}, replaces the outer average with a maximum,
\begin{equation}
  \bar{p}_s^{\,\text{xv}}
  \;=\;
  \max\bigl\{\bar{p}_s^{\,\text{CC}},\, \bar{p}_s^{\,\text{MLO}}\bigr\},
  \label{eq:max_views}
\end{equation}
corresponding to the heuristic that a patient should be flagged on the basis of her more suspicious view. The fourth rule, \textit{max image}, takes the maximum across all of the patient's images,
\begin{equation}
  \bar{p}_s^{\,\text{xi}}
  \;=\;
  \max_{i \in \mathcal{I}(s)} p_i,
  \label{eq:max_image}
\end{equation}
operationalizing the principle that a patient-level decision should reflect her single most suspicious image. The use of patient-level
(or exam-level) rather than image-level predictions in multi-view mammography classification was established by~\citet{carneiro_unregistered_2017}, who explicitly framed the classification target as the whole mammographic exam containing both CC and MLO views rather than the individual lesion. The
\textit{max image} rule is the simplest aggregator consistent with that framing and admits a probabilistic interpretation as a noisy-OR combination of the per-image probabilities~\citep{kraus_yao_andriluka_2016}: if each image provides an independent noisy observation of the underlying patient-level event, the maximum is the maximum-likelihood combination under the assumption that at most one positive image is needed for a positive patient.

The four rules span the natural axis from full averaging (\textit{raw mean}, equal weight on every image) to extreme order-statistic selection (\textit{max image}, using only the maximum), with the intermediate \textit{mean views} and \textit{max views} testing whether view structure carries information beyond the within-view average. For the normal-versus-malignant task, the patient-level label follows the any-positive rule of Section~\ref{sec:patient_grouping}: a patient is labeled malignant if any of her retained images is malignant, and normal otherwise (the per-task form of the rule is given in Section~\ref{sec:patient_grouping}). This rule matches the clinical target, detection of malignancy in at least one breast, and is the natural label partner for \textit{max image} fusion, which flags a
patient on her single most suspicious image; scoring an any-positive predictor against a majority-vote label would mismatch the aggregation rule and the target it is meant to estimate. Patient-level AUC is reported alongside the image-level AUC for each fusion rule. The choice of fusion rule is a methodological one rather than a heuristic,
and the four rules are compared empirically in
Section~\ref{sec:res-fusion}, since the same image-level predictions can yield noticeably different patient-level AUCs depending on how they
are combined.

\subsection{Patient-cluster bootstrap}
\label{sec:bootstrap}

The cross-validation scheme of Section~\ref{sec:cv} provides point estimates of patient-level AUC, but not the uncertainty around those estimates. Patient-grouped cross-validation controls bias in the point estimate by enforcing leakage-free partitions;
the patient-cluster bootstrap described in this subsection is its inferential complement, attributing the variance around the point
estimate to the right unit of statistical replication. The combination of the two --- CV for bias control, cluster bootstrap for variance attribution --- is what makes the AUC differences
reported in Sections~\ref{sec:res-modality}---\ref{sec:res-resnet} interpretable as statements about the patient population rather than about a particular fold structure or image collection.

The standard nonparametric bootstrap of
\citet{efron_1979_bootstrap} resamples observations independently and is not directly applicable here. The unit of statistical replication in our analysis is the patient rather than the image, and the images contributed by a single patient are correlated:
they share anatomy, acquisition geometry, and pathology. We therefore use the patient-cluster bootstrap, in which the resampling unit is the patient rather than the  image~\citep{field_welsh_2007}. For each reported AUC we draw $B = 2{,}000$ bootstrap resamples with replacement at the patient level, with each resample retaining every image contributed by the drawn patients so that the within-patient correlation structure of the original sample is preserved exactly. The patient-level AUC is recomputed on each resampled set, and the $2.5$th and $97.5$th empirical percentiles of the resulting bootstrap distribution give the $95\%$ confidence interval reported in the results. When AUCs from competing models are compared, for example CM-only versus
DM-only or one fusion rule versus another, the same $B$ patient-level resamples are reused for both models, so that the two models are evaluated on the same simulated patient population at every iteration and the AUC difference is taken within each resample. This is the bootstrap analogue of a paired comparison: any variability driven by which patients happen to appear in a given resample cancels out of the difference rather than inflating it.

The patient-cluster bootstrap has two practical advantages over alternative inference procedures for AUC. First, unlike the DeLong test~\citep{delong_1988}, it does not rely on the asymptotic normality of the AUC statistic, and, more importantly, it does not assume
independent observations --- an assumption that a clustered design violates, since each patient contributes several correlated images. Applied to image-level AUCs, the DeLong procedure would understate uncertainty for the same within-patient correlation reason that motivates the patient-grouped cross-validation of Section~\ref{sec:cv}. Second, unlike a standard non-clustered bootstrap, which resamples images independently and so breaks the within-patient correlation structure, the patient-cluster bootstrap resamples whole patients and retains every image of each drawn patient, so the uncertainty it reports is attached to the patient --- the level at which the model must generalize --- rather than to the image.

\section{Results}
\label{sec:results}
\subsection{Patient-grouped baseline}
\label{sec:res-baseline}

\begin{figure}[!htbp]
\centering
\includegraphics[width=0.95\linewidth]{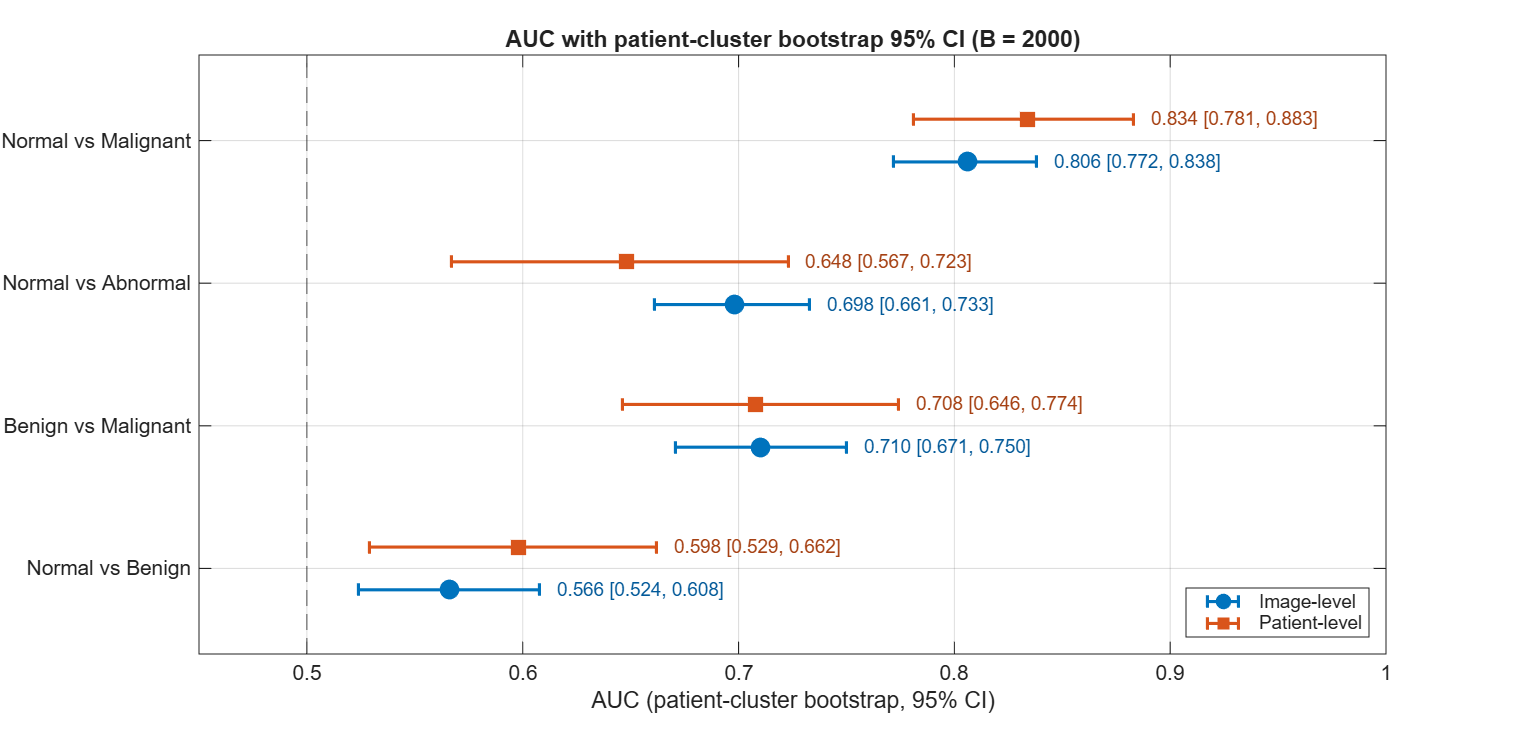}
\caption{Patient-cluster bootstrap 95\% confidence intervals for image- and patient-level AUC across the four binary diagnostic tasks under the DM+CM combined model. The normal-versus-malignant task is the strongest discriminator, with both image- and patient-level confidence intervals well separated from chance; the normal-versus-benign and normal-versus-abnormal tasks show wider patient-level intervals whose lower bounds approach $0.53$.}
\label{fig:forest_main}
\end{figure}

We begin by evaluating four binary diagnostic tasks --- normal versus malignant, normal versus abnormal (benign or malignant), benign versus malignant, and normal versus benign --- under the patient-grouped
framework of Section~\ref{sec:cv}, applied to the pooled DM+CM training set, both image types as rows with a binary contrast-type indicator. For each task, we report the image-level AUC computed from pooled out-of-fold probabilities, the patient-level AUC obtained from \textit{raw mean} fusion across each patient's images, and 95\% confidence intervals constructed by the patient-cluster bootstrap procedure described in Section~\ref{sec:bootstrap}. The complete set of point estimates and confidence intervals is collected in Table~\ref{tab:baseline} and displayed graphically in Figure~\ref{fig:forest_main}.

\begin{table}[!htbp]
\centering
\caption{Patient-grouped nested cross-validation results for the four binary tasks, DM+CM combined model. Per-image AUC is computed from pooled out-of-fold probabilities; patient-level AUC uses \textit{raw mean} fusion. Confidence intervals are 2000 patient-cluster bootstrap percentile intervals.}
\label{tab:baseline}
\small
\begin{tabular}{lcc}
\toprule
Task & Image AUC (95\% CI) & Patient AUC (95\% CI) \\
\midrule
Normal vs.\ Malignant   & 0.806 (0.772--0.838) & 0.834 (0.781--0.883) \\
Normal vs.\ Abnormal    & 0.698 (0.661--0.733) & 0.648 (0.567--0.723) \\
Benign vs.\ Malignant   & 0.710 (0.671--0.750) & 0.708 (0.646--0.774) \\
Normal vs.\ Benign      & 0.566 (0.524--0.608) & 0.598 (0.529--0.662) \\
\bottomrule
\end{tabular}
\end{table}

The clearest pattern in Table~\ref{tab:baseline} and
Figure~\ref{fig:forest_main} is the ordering of the four tasks by separability. Normal-versus-malignant discrimination is substantially the strongest, with patient-level AUC $0.834$ ($95\%~\text{CI}~0.781$--$0.883$) and an image-level AUC of $0.806$
($95\%~\text{CI}~0.772$--$0.838$). Benign-versus-malignant is the next-strongest contrast, with patient-level AUC $0.708$ ($95\%~\text{CI}~0.646$--$0.774$), separated from normal-versus-malignant but with overlapping intervals. The normal-versus-abnormal task ($0.648$, $0.567$--$0.723$) and the
normal-versus-benign task ($0.598$, $0.529$--$0.662$) are weaker, with wide patient-level intervals whose lower bounds approach $0.52$. This
ordering is consistent with the radiological intuition that contrast uptake on CESM is most informative for distinguishing healthy tissue
from frank malignancy, while benign lesions can resemble either malignancies or normal tissue depending on their composition.

The point estimates in Table~\ref{tab:baseline} are stable to the choice of cross-validation partition: image-level AUC standard deviations across the ten outer-CV repeats lie between $0.003$ and $0.006$ for every task, and the bootstrap intervals, computed under a separate form of resampling, are correspondingly narrow.

In several tasks the patient-level AUC exceeds the image-level AUC (for normal-versus-malignant, $0.834$ versus $0.806$), which may seem
counterintuitive but follows directly from how the two quantities are defined. Every image inherits its breast-level label, so a malignant patient contributes both the few images on which the malignancy is visible and the several uninformative images --- the contralateral breast and unaffected views --- that look essentially normal and are
scored low by the classifier. At the image level these uninformative images are labeled malignant yet ranked low, and they count against  the image-level AUC. Aggregating a patient's images into a single
score by \textit{raw mean} averages out this within-patient noise: the informative images pull the patient's mean up enough to rank her correctly, and the uninformative images are absorbed into one
per-patient value rather than counted as individual ranking errors. The patient-level estimate is therefore the average of several noisy per-image predictions, which is a lower-variance estimate of the quantity of interest than any single image provides. This image-to-patient improvement is distinct from, and in the opposite direction to, the image-level \emph{inflation} produced by information leakage (Section~\ref{sec:intro}): the inflation arises when a patient's images straddle the train/test split so the model recognizes the patient rather than the disease, whereas the improvement described here arises under a strictly patient-grouped partition, where no such leakage is possible and the only effect of aggregation is to reduce per-patient variance.

The gap between the patient-grouped and image-level cross-validation estimates on the same data is the more substantive check. For the normal-versus-malignant task, image-level five-fold cross-validation (Supplement~S4) yields an image-level AUC of $0.818$, only $0.012$ above the patient-grouped image-level estimate of
$0.806$. A feature bank that had memorized patient-specific structure rather than learning generalizable signal would lose far more performance when patient identity is removed from the partitioning. 

Rather than inheriting the large worst-case inflation reported in settings such as slice-level brain MRI (Section~\ref{sec:intro}), the feature bank shows almost no image-to-patient gap here, consistent with patients contributing roughly eight moderately correlated images rather than dozens of near-identical slices. Its near-absence is reassurance that the modality, fusion-rule, and classifier comparisons that follow reflect properties of the feature representation, not leakage between training and test partitions.

\subsubsection{Discrimination floor on benign-versus-malignant}
\label{sec:res-scope}

Benign-versus-malignant is the clinically meaningful next step beyond ``is there cancer or not,'' but its patient-level AUC of $0.708$ ($95\%~\text{CI}~0.646$--$0.774$) is markedly weaker than normal-versus-malignant. The question is whether this shortfall is a limitation of the wavelet feature bank or of the contrast itself at this sample size. Two findings indicate the latter.

First, a pretrained ResNet-50 frozen-feature representation, run on the same patients under identical patient-grouped folds, does no better.
Under \textit{max image} fusion the wavelet bank reaches $0.713$ ($0.650$--$0.778$) on CM and $0.690$ ($0.625$--$0.753$) on DM+CM, and ResNet-50 reaches $0.746$ ($0.687$--$0.806$) and $0.710$ ($0.645$--$0.770$) on the same patients; all four intervals overlap substantially and the two representations are within about $0.04$ AUC everywhere (full results in Supplement~S3). A much stronger representation does not rescue the task.

Second, the \textit{max image} fusion lift that helps
normal-versus-malignant (Section~\ref{sec:res-fusion}) disappears here: across the four representation--modality cells the four fusion rules
fall within $\pm 0.02$--$0.04$ of one another, and for the wavelet DM+CM cell \textit{max image} is in fact the lowest. This is what one expects when benign and malignant lesions overlap in their CESM
enhancement far more than either does with normal tissue; no single image is decisively more suspicious, so flagging on the most suspicious image buys nothing.

The benign-versus-malignant contrast on CDD-CESM is therefore task-limited, not representation-limited. The remaining analyses focus on normal-versus-malignant, the most clinically actionable task and the one on which the dataset supports the most informative inference.

\subsection{Modality comparison: DM, CM, and DM + CM}
\label{sec:res-modality}

\begin{figure}[!htbp]
\centering
\includegraphics[width=\linewidth]{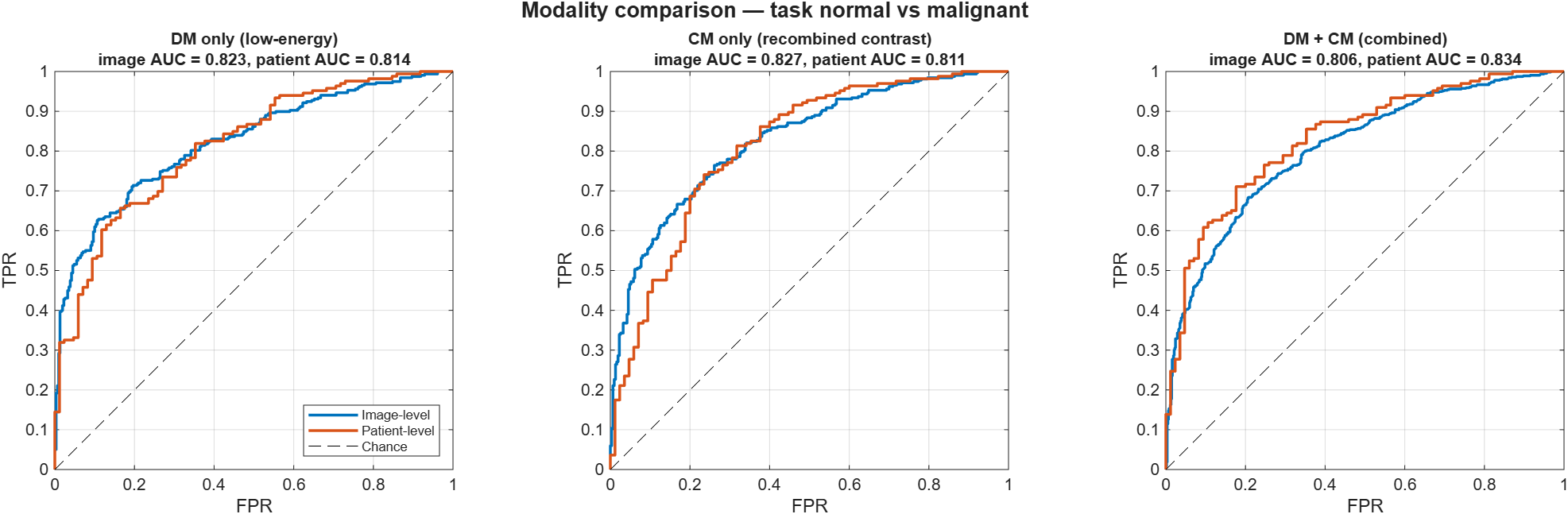}
\caption{ROC curves for the normal-versus-malignant task under the three modality configurations. Image-level curves (blue) are computed from pooled out-of-fold probabilities; patient-level curves (red) use
\textit{raw mean} fusion. The three configurations are closely comparable, with image-level AUCs of $0.823$ (DM), $0.827$ (CM), and $0.806$ (DM+CM), and patient-level \textit{raw mean} AUCs clustered near $0.81$--$0.83$.}
\label{fig:modality_roc}
\end{figure}

\begin{figure}[!htbp]
\centering
\includegraphics[width=\linewidth]{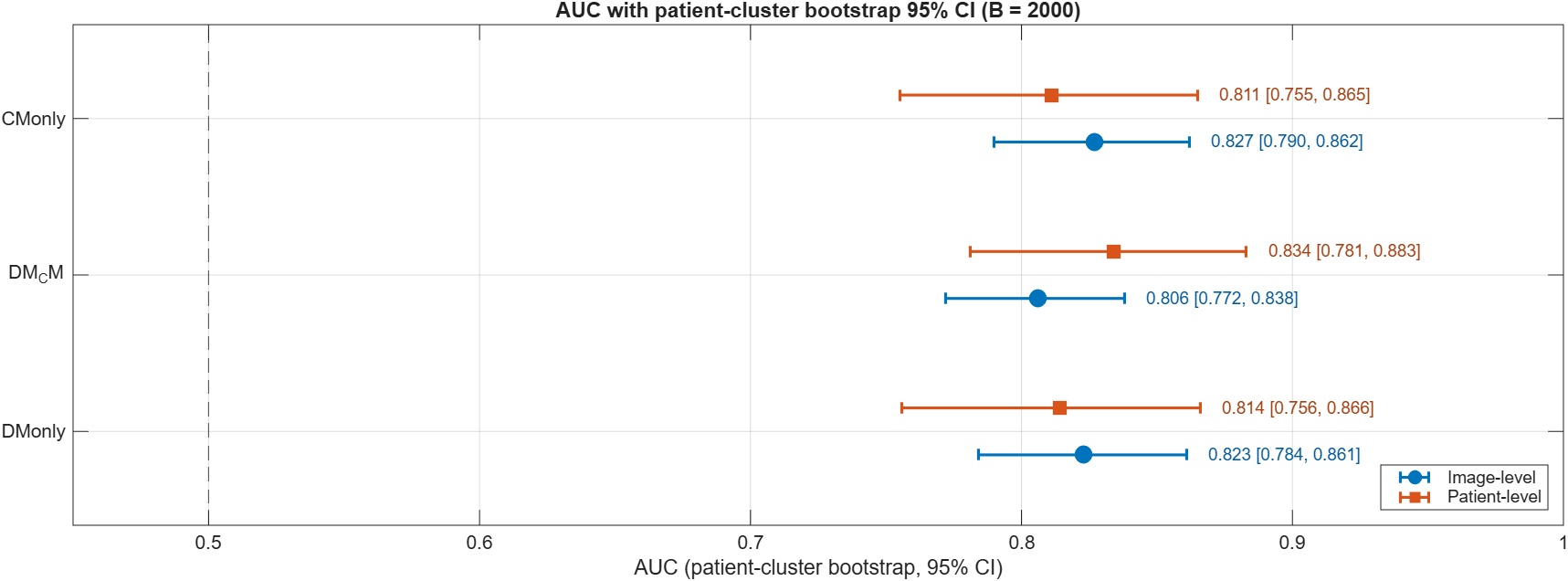}
\caption{Patient-cluster bootstrap 95\% confidence intervals for image- and patient-level AUC on the normal-versus-malignant task, separately for the three modality configurations. The three configurations overlap substantially at both evaluation scales; no modality separates from the others beyond bootstrap noise, although DM+CM is highest at the patient level by point estimate.}
\label{fig:modality_forest}
\end{figure}

A central methodological question raised in Section~\ref{sec:intro} is how much of the malignancy-relevant signal lives in each of the two
acquired CESM image types when the same masked complex NDWT feature bank is applied to each. To answer it, we re-ran the normal-versus-malignant pipeline separately on the low-energy images (DM only), the recombined contrast images (CM only), and the pooled set of both image types (DM+CM). In the DM+CM configuration the DM and CM images enter as separate training rows rather than being concatenated
into a joint feature vector, so the predictor count is the 203 wavelet descriptors used by each single-modality model plus one binary contrast-type indicator distinguishing the two acquisitions, 204 predictors in total. The patient-grouped folds and bootstrap resamples were held fixed across the three configurations so that the differences in AUC are attributable to the choice of modality alone. Image- and patient-level ROC curves for each configuration are shown in Figure~\ref{fig:modality_roc}, and the corresponding AUCs with 95\% patient-cluster bootstrap confidence intervals are displayed in
Figure~\ref{fig:modality_forest}.

Two findings stand out. First, the three modality configurations are statistically indistinguishable on the normal-versus-malignant task. At the image level the AUCs are $0.823$ (DM), $0.827$ (CM), and $0.806$ (DM+CM); at the patient level under \textit{raw mean} fusion they are
$0.814$ (DM, $95\%~\text{CI}~0.756$--$0.866$), $0.811$ (CM, $0.755$--$0.865$), and $0.834$ (DM+CM, $0.781$--$0.883$), with every pairwise bootstrap interval overlapping substantially. The recombined
contrast image and the low-energy image therefore carry comparable malignancy-relevant signal when the same masked complex NDWT feature bank is applied to each. This runs counter to the expectation that the
contrast image, engineered to suppress background tissue and highlight iodine uptake, should dominate: the low-energy image is an equally informative substrate for this feature bank. The two acquisitions
nonetheless recover that comparable signal through largely disjoint, physically interpretable feature families (Section~\ref{sec:res-features}), so the equivalence in accuracy does not imply redundancy in mechanism.

Second, pooling DM and CM is the strongest configuration at the patient level, though by a margin within bootstrap noise rather than a decisive one. The DM+CM patient-level AUC of $0.834$ (\textit{raw mean}) and $0.883$ (\textit{max image}, Section~\ref{sec:res-fusion}) edges both single-modality models at every fusion rule, while its image-level AUC ($0.806$) is marginally below the two single-modality models. Because the pooled model fits a single shared coefficient vector across both image types and draws on the disjoint feature families each modality
supplies, the combined configuration loses nothing relative to either alone and is modestly stronger once predictions are aggregated to the patient level. From a deployment standpoint, a single-modality CM-only or DM-only pipeline remains a defensible simplification, since neither is distinguishable from the combined model at the patient level on this dataset.

The image-level versus patient-level gap also differs across configurations. The single-modality models and the combined model move differently from image-level to patient-level evaluation: combining the two modalities doubles the number of images per patient that enter the \textit{raw mean} fusion, and averaging across a larger image set reduces per-patient variance in the fused probability. The gap, in other words, is sensitive to the number of images being averaged, not only to the underlying classifier. We return to this point in Section~\ref{sec:res-fusion}, where the fusion-rule comparison shows that the gap is also strongly modulated by which images are aggregated.

\subsection{Stable feature selection across modalities}
\label{sec:res-features}
\begin{figure}[!htbp]
\centering
\includegraphics[width=\linewidth]{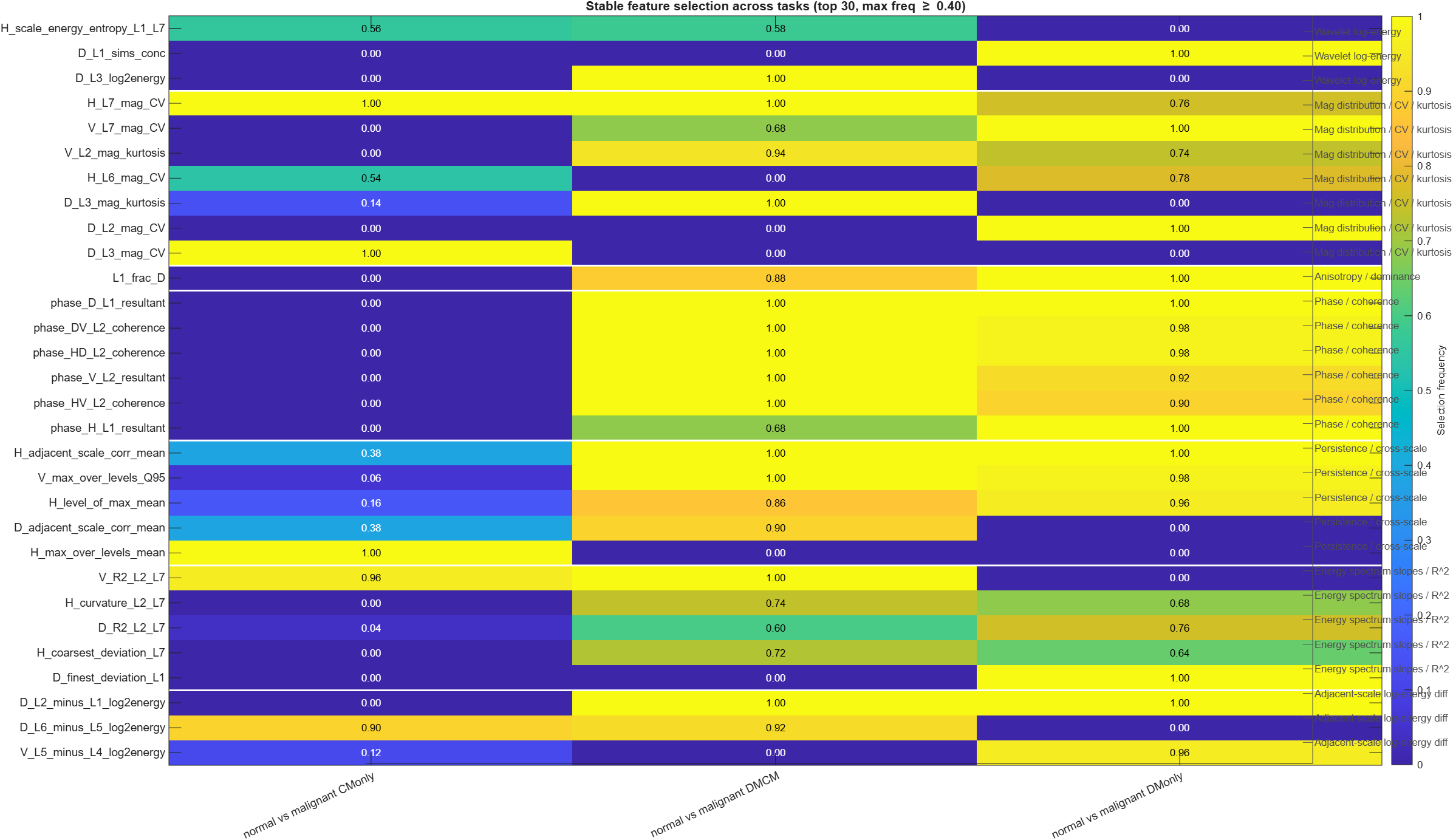}
\caption{Selection frequency of the top thirty stably selected features across outer-fold fits of the normal-versus-malignant pipeline, shown separately for the CM-only, DM-only, and DM+CM
configurations. Color encodes the fraction of outer-fold fits in which each feature has a non-zero elastic-net coefficient. The right margin labels each feature by its descriptor family. Phase-coherence and phase-resultant descriptors are selected in nearly every DM-only fit and almost never in the CM-only model, while the CM-only model
concentrates on a set of magnitude and log-energy descriptors.}
\label{fig:feature_heatmap}
\end{figure}

The modality comparison of Section~\ref{sec:res-modality} establishes that the DM-only and CM-only configurations reach statistically indistinguishable patient-level AUC on the normal-versus-malignant task. This raises a mechanistic question: do the two image types reach comparable accuracy by drawing on the same descriptors, or by different routes through the feature bank? To address it, we recorded the
elastic-net selection frequency of every feature across all outer-fold fits of the normal-versus-malignant pipeline, separately for the three
modality configurations. A feature is considered \emph{stably selected} if it appears in the active set of at least $50\%$ of the fitted models; the top thirty stably selected features across the three
configurations are displayed in Figure~\ref{fig:feature_heatmap}.

The pattern in Figure~\ref{fig:feature_heatmap} is the central mechanistic finding of the paper. The DM-only and CM-only models, applied to images of the same patients with the same masked complex
NDWT feature bank, reach equivalent accuracy through largely disjoint feature sets. The phase-coherence and phase-resultant descriptors are the clearest case. The most prominent examples are the level-1 and level-2 phase-resultant and
inter-channel phase-coherence descriptors, which are selected in nearly every DM-only
outer-fold fit and essentially never in the CM-only fits. The persistence and anisotropy descriptors show the same asymmetry: features such as \texttt{L1\_frac\_D},
\texttt{V\_max\_over\_levels\_Q95}, and
\texttt{H\_adjacent\_scale\_corr\_mean} are selected by nearly every DM-only fit and rarely by the CM-only model, as are several adjacent-scale log-energy differences such as \texttt{V\_L5\_minus\_L4\_log2energy} and \texttt{D\_L2\_minus\_L1\_log2energy} (though the direction is not uniform across this family --- \texttt{D\_L6\_minus\_L5\_log2energy}, for instance, is selected predominantly by the CM-only model). The
CM-only model in turn concentrates on a largely separate set: magnitude coefficient-of-variation and persistence descriptors such as \texttt{D\_L3\_mag\_CV} and \texttt{H\_max\_over\_levels\_mean} (both at $100\%$ selection) and the energy-spectrum $R^2$ feature
\texttt{V\_R2\_L2\_L7} ($96\%$), each of which is rarely or never selected by the DM-only model. A smaller number of magnitude descriptors are selected by both models --- chiefly broadband coefficient-of-variation features such as \texttt{H\_L7\_mag\_CV}
(selected in essentially every CM-only fit and in roughly three-quarters of DM-only fits) and \texttt{H\_L6\_mag\_CV} --- so the disjointness is
specific to the phase channel and the geometric descriptors rather than total. The DM+CM column shows a hybrid pattern: features from both
groups appear, reflecting that the combined model has access to both signal carriers and the elastic-net penalty distributes weight between them.

The disjointness of the selected feature sets has a physical interpretation that explains how the DM-only and CM-only models reach comparable patient-level AUC while sharing few descriptors. The DM
image is a conventional mammogram that depicts overall tissue density and architectural organization; the signal that distinguishes
malignant from normal tissue in this image is geometric, such as distortion of fibroglandular patterns, oriented texture irregularities, and disrupted local coherence. It is best captured by features that summarize the phase of complex wavelet coefficients, since phase coherence indexes how consistently local oriented structure is organized within a scale. The CM image, in contrast, suppresses background tissue and selectively highlights iodine uptake; the malignancy signal here is intensity-localized rather than geometric, and is best captured by features that summarize the magnitude distribution of the wavelet coefficients; in particular the coefficient of variation, which is large when energy is concentrated in a small fraction of the tissue support. The masked complex NDWT framework adapts to whichever signal carrier is dominant in a given modality, drawing on its phase summaries when the image is geometric and on its magnitude summaries when the image is intensity-localized. This adaptive behavior is not a property of the elastic-net classifier in isolation; it is a property of the feature bank, which provides both
phase and magnitude descriptors at every scale and direction and lets the classifier select among them. The two acquisitions thus encode comparable malignancy information through different physical channels, and the feature bank is expressive enough to read both, which is why the modality equivalence of Section~\ref{sec:res-modality} coexists with near-disjoint feature selection rather than implying redundancy between the two image types.

\subsection{Patient-level fusion}
\label{sec:res-fusion}

\begin{figure}[!htbp]
\centering
\includegraphics[width=0.95\linewidth]{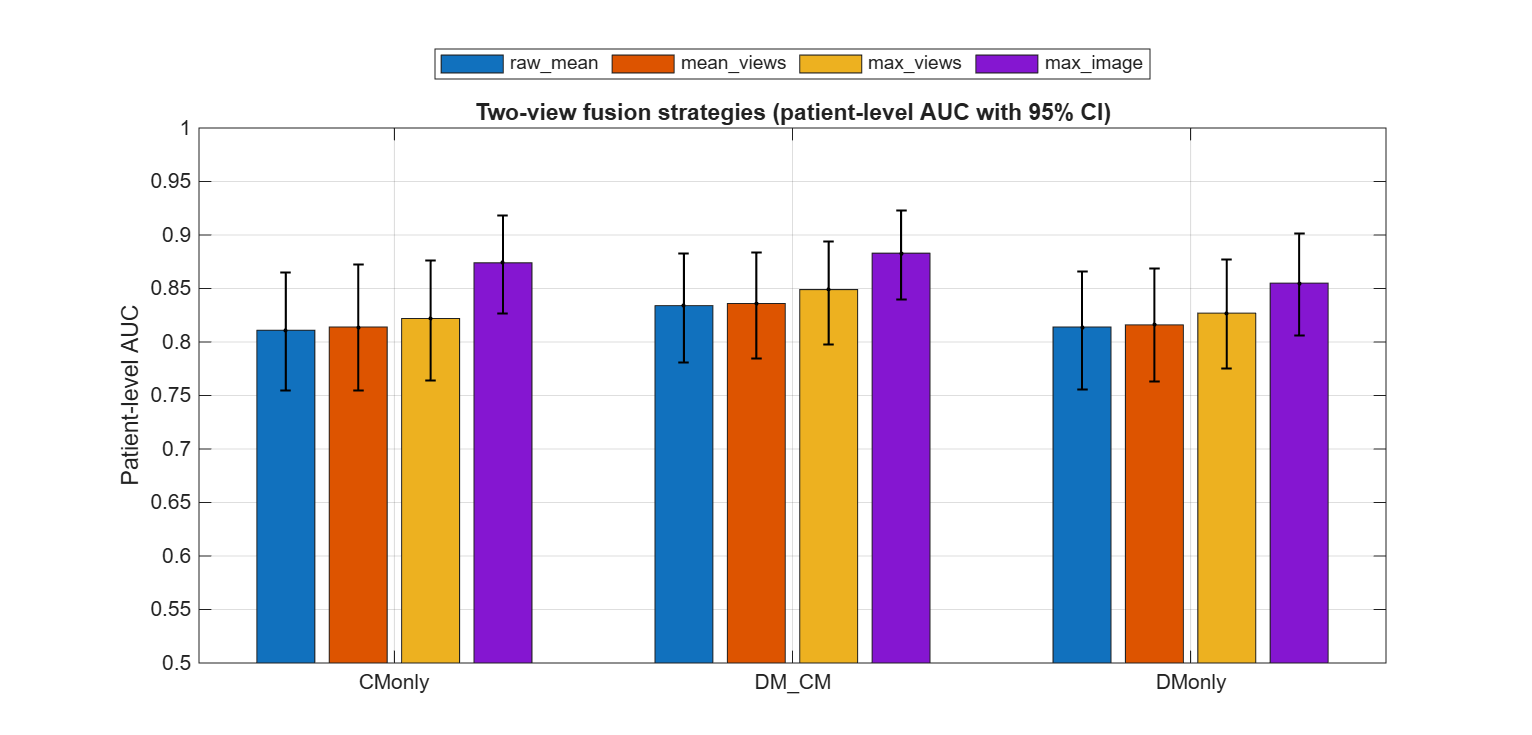}
\caption{Patient-level AUC under four fusion rules, for each of the three modality configurations. The \textit{raw mean}, \textit{mean views}, and \textit{max views} rules give nearly identical patient-level AUCs within each configuration; the \textit{max image} rule lifts the patient-level AUC by roughly $0.04$--$0.07$ in every configuration. Error bars are $95\%$ patient-cluster bootstrap percentile intervals.}
\label{fig:fusion}
\end{figure}

We next ask how the choice of fusion rule (Section~\ref{sec:fusion}) affects the patient-level AUC, holding the underlying image-level
predictions fixed. The four rules --- \textit{raw mean}, \textit{mean views}, \textit{max views}, and \textit{max image} --- were applied to the same out-of-fold image probabilities used in Section~\ref{sec:res-modality}, separately for each of the three modality configurations. The resulting patient-level AUCs and 95\% bootstrap confidence intervals are displayed in Figure~\ref{fig:fusion}.

The pattern in Figure~\ref{fig:fusion} divides the four rules into two groups. The three averaging-style rules --- \textit{raw mean}, \textit{mean views}, and \textit{max views} --- behave similarly within each modality, indicating that how the
within-view averaging is organized (whether views are weighted equally or compared at the view level) matters relatively little. What does matter is whether the patient-level summary is an average
at all. The \textit{max image} rule, which abandons averaging entirely and reports the patient on her single most suspicious image, separates from the other three in every modality configuration, with the largest gain over \textit{raw mean} on the
single-modality models. The size of the lift (roughly
0.04--0.07  patient-level AUC across modalities) is reproduced when the feature bank is changed to ResNet-50 (Section~\ref{sec:res-resnet}), so it is a property of the fusion rule rather than of the particular feature representation.

Averaging dilutes the signal for a simple clinical reason. Of a patient's eight or so images, only the two or three views of the affected breast carry visible malignancy; the contralateral breast and unaffected views look essentially normal and are scored low. Averaging pulls the high probability of the informative images toward these low ones, so the patient-level estimate mixes evidence with non-evidence. The maximum instead returns the single most confident image, letting the patient-level summary reflect the strongest finding rather than the average across the study. 

This is the same logic that radiologists apply when assigning a final BI-RADS category from a multi-view mammographic exam: the overall assessment reflects the most suspicious finding, not the
average across views~\citep{ACRBI-RADS2013, zonderland_birads_2014}. Applied as an aggregation rule on classifier probabilities, the same
heuristic gives better patient-level discrimination than averaging. Under \textit{max image} fusion, the CM-only model achieves a patient-level AUC of $0.874$ ($95\%~\text{CI}~0.827$--$0.918$) and the DM+CM model achieves $0.883$ ($95\%~\text{CI}~0.840$--$0.923$); the DM-only model reaches $0.855$ ($0.806$--$0.901$).

Three observations follow. The first is that the framework reaches patient-level AUC near $0.88$ on the most clinically actionable contrast in this dataset, using a model in which every retained predictor is a named, physically interpretable wavelet descriptor. The classifier is small, fast to train, and transparent at the level of named
wavelet-feature groups and selected predictors.

The second is that the three modality configurations remain close under \textit{max image} fusion, as under the averaging rules (Section~\ref{sec:res-modality}): DM, CM, and DM+CM reach $0.855$, $0.874$, and $0.883$ respectively, with overlapping bootstrap intervals. The combined configuration is highest by point estimate, but a single-modality pipeline incurs no statistically discernible
loss, so a classification pipeline can be built from either acquisition alone with no measurable cost in patient-level performance.

The third is that how the per-image predictions are combined into a patient-level prediction matters more than is usually acknowledged. The gap between averaging across images and taking the most suspicious image is on the order of $0.04$--$0.07$ AUC, large enough that two papers using the same underlying classifier can report substantially different patient-level performance just by choosing
different fusion rules. Fusion choice should therefore be reported explicitly in any CESM classifier evaluation, alongside the classifier itself.

\subsection{Robustness to classifier choice}
\label{sec:res-classifier}

\begin{figure}[!htbp]
\centering
\includegraphics[width=\linewidth]{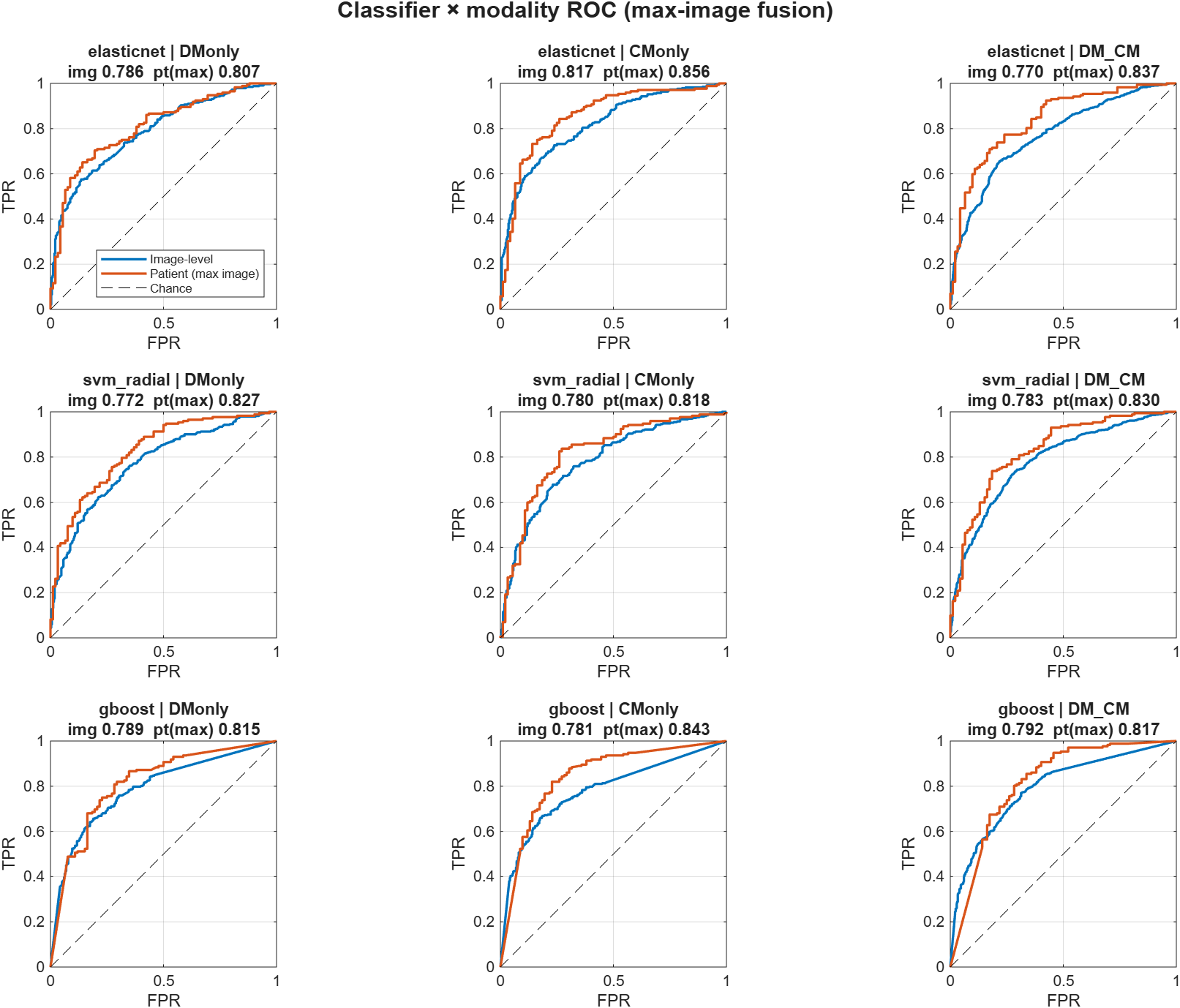}
\caption{Patient-level ROC curves for the normal-versus-malignant task under max image
fusion, stratified by classifier and modality. The panels show that modality ordering is not
stable across classifier families, consistent with the bootstrap interval results.}
\label{fig:classifier_compare}
\end{figure}

The modality-equivalence finding is not an artifact of the linear elastic-net classifier. We re-ran the normal-versus-malignant pipeline with a radial-basis SVM and a gradient-boosted tree ensemble, holding the feature bank, patient-grouped folds, fusion rules, and bootstrap resamples fixed, and compared the three classifiers at the patient level under \textit{max image} fusion. The patient-level AUCs are displayed in Figure~\ref{fig:classifier_compare}.

Two patterns hold across all three classifiers. First, the elastic-net model is not statistically outperformed by either nonlinear classifier. On every modality the three classifiers fall within a narrow band whose bootstrap intervals overlap substantially, so no
classifier separates from the others by a margin the dataset can resolve: on DM the three patient-level AUCs are $0.807$ (elastic-net), $0.827$ (SVM), and $0.815$ (gradient boosting); on CM they are $0.856$,
$0.818$, and $0.843$; on DM+CM they are $0.837$, $0.830$, and $0.817$. The SVM and gradient-boosting estimates reach patient-level AUC of approximately $0.84$ at their best modality, neither exceeding the
elastic-net estimate beyond bootstrap noise. These figures are not directly comparable to the headline elastic-net AUCs of Sections~\ref{sec:res-fusion} and~\ref{sec:res-resnet}: to give the three classifiers a common tuning budget, the classifier comparison fixes the elastic-net mixing parameter at $\alpha = 0.5$ with a coarser regularization path, whereas the headline pipeline sweeps $\alpha$ over
$\{0.3, 0.5, 0.7\}$ on a finer path, so the elastic-net column here is internally consistent with the SVM and gradient-boosting columns rather than with the full-pipeline AUCs reported elsewhere. Second, the modality equivalence reported in Section~\ref{sec:res-modality} is reproduced under each classifier rather than overturned by it: no configuration is reliably
strongest across the three classifiers. The best modality is in fact different for each classifier --- DM+CM for the SVM, CM for gradient boosting, and CM for the elastic-net --- and in every case the
bootstrap intervals of the three modalities overlap, so the apparent ordering shifts with the classifier precisely because the three modalities are statistically indistinguishable. The absence of a
stable modality ranking across classifiers is exactly what the modality-equivalence finding predicts, and it confirms that the equivalence is a property of the masked complex NDWT feature bank and the patient-grouped evaluation rather than of the linear model.

The practical reading is that the elastic-net classifier, chosen for its sparsity, speed, and per-feature interpretability, sacrifices no measurable patient-level discrimination to two more flexible function classes on this feature representation. The conclusions of Sections~\ref{sec:res-modality} and~\ref{sec:res-fusion} are properties of the masked complex NDWT feature bank and the patient-grouped evaluation, not of the particular classifier used to read it.

\subsection{Comparison with a deep-learning feature representation}
\label{sec:res-resnet}

To benchmark the masked complex NDWT feature bank against a deep-learning representation, we substituted the $2{,}048$-dimensional
activations of the global average pooling layer of an ImageNet-pretrained ResNet-50~\citep{he_resnet_2016, deng_imagenet_2009}
for the $203$ wavelet features and ran the resulting design through the same patient-grouped pipeline (same folds, classifier, fusion
rules, and bootstrap). This frozen-feature protocol is the standard transfer-learning baseline when end-to-end fine-tuning is not
feasible~\citep{sharif_razavian_cnn_features_2014,
yosinski_transferable_2014}, and pits a purpose-built domain representation of $203$ predictors against a generic representation of ten times the dimensionality, pretrained on millions of natural
images.

\begin{figure}[!htbp]
\centering
\includegraphics[width=0.8\linewidth]{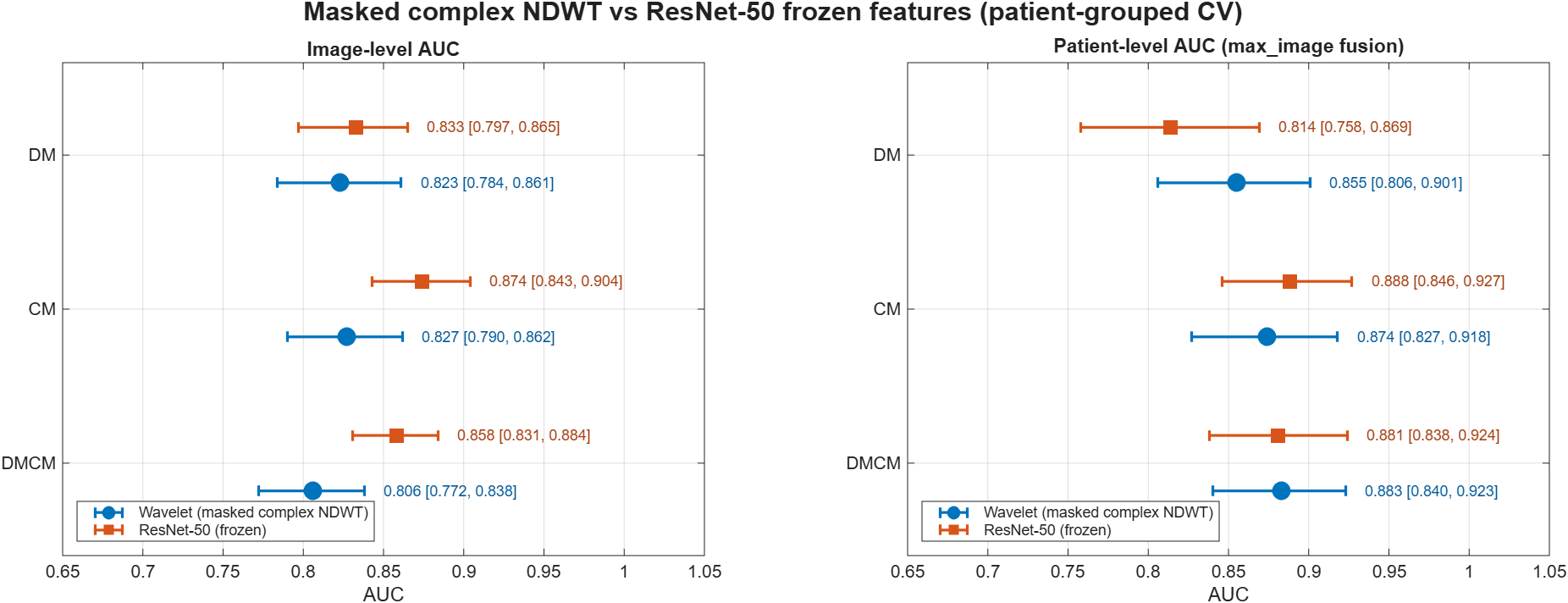}
\caption{Patient-level AUC under \textit{max image} fusion for the masked complex NDWT feature bank (blue circles) and the ResNet-50 frozen-feature baseline (orange squares), on the three modality configurations. Bootstrap 95\% CIs overlap substantially across all three configurations.}
\label{fig:resnet_compare_headline}
\end{figure}

Under \textit{max image} fusion, the wavelet and ResNet-50 representations reach patient-level AUCs whose bootstrap intervals overlap in all three modality configurations: wavelet $0.855$
($0.806$--$0.901$) versus ResNet $0.814$ ($0.758$--$0.869$) on DM, $0.874$ ($0.827$--$0.918$) versus $0.888$ ($0.846$--$0.927$) on CM, and
$0.883$ ($0.840$--$0.923$) versus $0.881$ ($0.838$--$0.924$) on DM+CM (Figure~\ref{fig:resnet_compare_headline}). The point estimates differ most on DM, where the wavelet bank is higher by $0.041$, and least on
DM+CM, where the two are within $0.002$; on CM, ResNet is higher by $0.014$. In none of the three configurations does the difference exceed
bootstrap noise, so the two representations are statistically indistinguishable at the patient level despite the deep representation having ten times the dimensionality.

\begin{figure}[!htbp]
\centering
\includegraphics[width=0.7\linewidth]{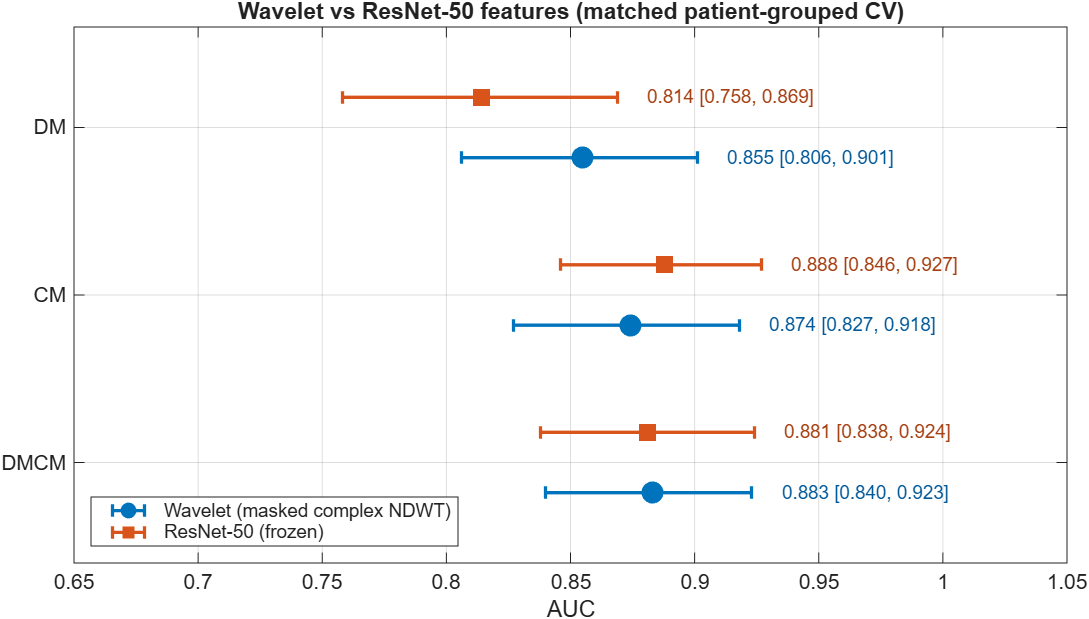}
\caption{Wavelet vs ResNet-50 frozen features under matched patient-grouped cross-validation. Left: image-level AUC. Right: patient-level AUC under \textit{max image} fusion. The image-level
edge of the ResNet representation narrows at the patient level. Error bars are 95\% patient-cluster bootstrap intervals ($B = 2000$).}
\label{fig:resnet_compare}
\end{figure}

At the image level the ResNet-50 representation is somewhat stronger than the wavelet bank (ResNet $0.833$, $0.874$, $0.858$ versus wavelet
$0.823$, $0.827$, $0.806$ on DM, CM, and DM+CM respectively). On CM and DM+CM this edge narrows under \textit{max image} fusion at the patient
level, where the two representations become statistically indistinguishable; on DM the patient-level ordering in fact reverses,
the wavelet bank moving from $0.010$ behind at the image level to $0.041$ ahead at the patient level, though with overlapping intervals at both scales. The ancillary fusion finding of the wavelet pipeline is also reproduced under the ResNet features: the \textit{max image} lift
over averaging-style fusion holds for the ResNet bank too, showing that the fusion-rule result of Section~\ref{sec:res-fusion} is not specific to the wavelet feature bank.

The parity itself is the substantive result, but its consequences for a radiologist using the model are what make it matter. The wavelet predictors map onto categories that are already used to read mammograms: oriented phase coherence indexes architectural distortion and spiculation, the dominant DM-image cues for malignancy in the
BI-RADS lexicon~\citep{ACRBI-RADS2013}; magnitude distribution on CM indexes localized iodine uptake, the dominant CM cue. A flagged case
can therefore be inspected at the level of specific scale-direction descriptors and traced back to image regions a radiologist could verify
against the imaging finding. This per-feature physical correspondence is the property that distinguishes the wavelet representation from a
generic deep representation of comparable patient-level accuracy, and the property that makes the parity result reported here a substantive
methodological contribution rather than a horse-race outcome.

The same qualitative parity pattern --- the wavelet and ResNet-50 representations producing overlapping patient-cluster bootstrap intervals at the patient level --- holds on the harder
benign-versus-malignant contrast (Section~\ref{sec:res-scope}, Supplement~S3), where both representations achieve substantially lower absolute performance but remain close to one another in every modality configuration. The parity finding is therefore not specific to the strongest contrast: under leakage-free evaluation on this dataset, the interpretable wavelet representation and the deep ResNet-50 representation reach comparable patient-level
accuracy whether the task is well-separated (normal versus malignant) or near the limit of what the dataset supports (benign versus malignant).

In short, the 203-dimensional interpretable wavelet bank shows no statistically discernible
patient-level AUC loss relative to a generic deep representation ten times its size, while retaining a feature selection profile that matches the categories radiologists already use to read CESM mammograms.

\section{Discussion}
\label{sec:discussion}

\paragraph{The two image types carry equivalent signal through different channels}

The central modality finding is one of equivalence. On the normal-versus-malignant task the DM-only, CM-only, and DM+CM configurations reach statistically indistinguishable patient-level AUC, with overlapping patient-cluster bootstrap intervals under every
fusion rule and under all three classifier families
(Sections~\ref{sec:res-modality} and~\ref{sec:res-classifier}). The recombined contrast image does not dominate the low-energy image, as the acquisition design might lead one to expect; the two acquisitions
are comparably informative inputs to the masked complex NDWT feature bank. This is a stronger and more surprising statement than a simple ranking of the two image types would be. The CM image is engineered to suppress background tissue and isolate iodine uptake, and the natural expectation is that a malignancy classifier should read it more accurately than the conventional low-energy image; instead, the
low-energy image carries an equal amount of malignancy-relevant signal once the feature bank is allowed to summarize both the magnitude and the phase of its wavelet coefficients. Pooling the two image types is highest at the patient level by point estimate but not by a margin the dataset can resolve, so a single-modality pipeline, built from either
acquisition alone, is a defensible simplification that loses no measurable patient-level performance. The same equivalence holds for the ResNet-50 frozen-feature representation (Section~\ref{sec:res-resnet}), so it is a property of the dataset and the patient-grouped evaluation rather than an artifact of the wavelet representation.

A practical consequence is that the common deep-learning default of using both acquisitions together is not required for this task at this dataset size: either image type alone supports comparable patient-level discrimination, and the choice between them can be made on grounds of acquisition cost or dose rather than expected accuracy.

\paragraph{Equivalent accuracy is reached through disjoint, physically interpretable feature channels}

The equivalence in accuracy does not imply redundancy in mechanism. The DM-only and CM-only models reach their comparable patient-level AUC through largely disjoint feature sets (Section~\ref{sec:res-features}, Figure~\ref{fig:feature_heatmap}). The
DM-only model is essentially a phase model: the top phase-resultant and phase-coherence features are selected in $90$--$100\%$ of DM-only outer folds and in essentially none of the CM-only folds. The CM-only model in turn concentrates on magnitude-distribution and energy-spectrum descriptors, chiefly coefficient-of-variation features, that the DM-only model rarely selects. The two acquisitions therefore encode
comparable malignancy information through different physical channels: geometric, oriented structure on the low-energy image, read by the phase channel of the complex transform, and intensity-localized iodine
uptake on the contrast image, read by the magnitude channel. 

To our knowledge, this is among the first demonstrations that complex-wavelet phase
summaries carry substantial mammographic malignancy signal under leakage-free,
patient-grouped evaluation, and it appears to be the first such demonstration in the
CDD-CESM setting.
 The pattern is consistent with the radiological intuition that malignancy detection on the low-energy image relies heavily on architectural disruption,
including spiculation, distortion, and asymmetric oriented patterns, which is encoded in the relative phase of oriented wavelet coefficients rather than in their magnitudes. A real-valued wavelet transform would
not give the classifier access to the phase channel at all, and a magnitude-only feature bank derived from a complex transform would discard it after the fact. The complex non-decimated wavelet transform preserves both magnitude and phase across scales, and the classifier
draws on whichever is more informative for the image type at hand; it is precisely this dual availability that allows two physically different acquisitions to be read with equal accuracy by a single feature bank. The role of complex wavelets in capturing oriented
texture has a long methodological history in statistical signal processing~\citep{vidakovic_book, kingsbury_complex_2001}; the contribution of the present analysis is to show that the phase channel
of such a transform, when summarized over an anatomical mask and evaluated under patient-grouped cross-validation, carries detectable malignancy signal at a sample size of $308$ patients, and that it does
so on the very image type the contrast acquisition was designed to improve upon.

\paragraph{Fusion rule choice is consequential and underreported}

The gap between \textit{raw mean} and \textit{max image} fusion is around $0.04$--$0.07$ AUC and is reproduced under a different feature representation (ResNet-50, Section~\ref{sec:res-resnet}), so it is not
specific to the wavelet bank. This makes a methodological point that extends beyond the present feature bank: when image-level classifiers are deployed on patients with multiple acquisitions, the choice of fusion rule is consequential, and the rule most closely aligned with the radiological principle that the final assessment is
driven by the most suspicious finding, max image fusion, also gives the highest empirical
AUC in this task. Reports that pool multi image patient predictions by averaging, as much of the deep-learning CESM literature implicitly does, may understate achievable patient-level performance. We recommend that fusion rule be reported explicitly in any multi-image medical classification study, alongside the classifier, the cross-validation scheme, and the unit of statistical replication; absent that information, two papers using the same underlying
classifier can report substantially different patient-level performance figures without the difference being attributable to anything beyond aggregation choice. The lift is also notably contrast-dependent: on the
harder benign-versus-malignant task, where benign and malignant lesions overlap in their enhancement patterns far more than either does with normal tissue, the \textit{max image} advantage disappears  (Section~\ref{sec:res-scope}), so the benefit of single-image fusion is specific to tasks in which the malignancy signal is concentrated in a small number of a patient's images rather than spread across them.

\paragraph{Interpretable representations match deep representations
under correct evaluation}

Under matched patient-grouped evaluation the ResNet-50 frozen-feature representation produces higher image-level AUC than the wavelet feature bank, with a clear margin on DM+CM, but this edge disappears at the
patient level under \textit{max image} fusion: across all three modality configurations the bootstrap intervals for the two representations overlap substantially. The pattern is informative in two ways. First, it indicates that part of the image-level advantage of the deeper representation reflects a finer-grained per-image signal that does not translate into a patient-level advantage once images are
aggregated by the maximum; within-patient variation in image quality, view, and contralateral imaging carries less weight under \textit{max image} than under averaging-style fusion. Second, it sharpens the interpretable-versus-opaque tradeoff for this task: at the clinically relevant level of aggregation, the $203$-dimensional wavelet bank gives up no measurable AUC to a $2{,}048$-dimensional ImageNet-pretrained representation, and recovers in exchange a  feature-selection profile in which every retained predictor has a named physical meaning. The same qualitative parity holds on the harder benign-versus-malignant contrast (Section~\ref{sec:res-scope}), at substantially lower absolute performance for both representations, so the parity finding generalizes across the range of task difficulty this dataset supports.

\paragraph{Limitations}

The CDD-CESM dataset is single-source and modest in scale; the feature-selection patterns and modality comparisons reported here should be reassessed on an independent CESM cohort when one becomes available. Each image's breast label is treated as the ground truth for the image, which is correct for contralateral comparisons but cannot capture lesion location or extent at the image level. The patient-cluster bootstrap uses percentile intervals. Bias-corrected and accelerated
cluster-bootstrap intervals could also be considered, especially if future studies use smaller
or more imbalanced cohorts; we did not rely on them here because the reported conclusions
are based mainly on paired bootstrap overlap and qualitative stability across configurations. The elastic-net classifier is linear in the
wavelet features; nonlinear interactions between features may exist that the linear model cannot capture, though the classifier-comparison
results in Section~\ref{sec:res-classifier} suggest that any such interactions are not substantial for this representation. Finally, the deep-learning comparison in Section~\ref{sec:res-resnet} uses a
frozen-feature transfer-learning baseline rather than an end-to-end fine-tuned network; this was a deliberate design choice given the dataset size, since fine-tuning a network of ResNet-50's capacity on
$308$ patients under patient-grouped cross-validation carries substantial overfitting risk. A fine-tuned configuration evaluated under the same patient-grouped folds is the natural extension of the present comparison and is left to future work.

\section{Conclusion}
\label{sec:conclusion}

We have built and evaluated a patient-level interpretable classification framework for contrast-enhanced spectral mammography, combining a masked complex non-decimated wavelet feature bank with an
elastic-net logistic classifier. Under repeated patient-grouped nested cross-validation with patient-cluster bootstrap confidence intervals on
the CDD-CESM dataset, the framework achieves patient-level AUC of $0.874$ ($95\%~\text{CI}~0.827$--$0.918$) for normal-versus-malignant detection using the recombined contrast image alone, $0.855$ ($0.806$--$0.901$) using the low-energy image alone, and $0.883$
($0.840$--$0.923$) using both image types together, all under single-image patient fusion and within bootstrap overlap of one another.

Four findings emerge from the analysis. First, the low-energy and recombined contrast images carry statistically indistinguishable malignancy-relevant signal for this feature bank, with no measurable patient-level advantage to either acquisition or to pooling the two; a single image type therefore suffices, and the choice between them can be made on grounds of acquisition cost or dose rather than expected accuracy. Second, the features the classifier selects on the two image types are nearly disjoint and physically interpretable: phase coherence on the low-energy image, where the malignancy signal is geometric, and magnitude distribution on the contrast image, where it is intensity-localized, so the two acquisitions reach equivalent accuracy through different physical channels rather than through redundant information. Third, single-image patient fusion, flagging on the most suspicious image rather than averaging across images, contributes a substantial patient-level lift on the normal-versus-malignant task that holds under a different feature representation as well, large enough that the choice of fusion rule should be reported alongside the
classifier in any multi-image medical classification study. Fourth, under matched patient-grouped evaluation, a pretrained ResNet-50 frozen-feature representation provides no statistically discernible
patient-level advantage over the interpretable wavelet feature bank on either the normal-versus-malignant or the benign-versus-malignant contrast, indicating that the deep representation offers no measurable AUC gain at the clinically relevant level of aggregation while
sacrificing the per-feature physical interpretability that the wavelet representation retains.

The framework matches nonlinear classifiers and a deep-learning feature representation at the patient level on the same patient-grouped folds, while remaining small, fast to train, and interpretable at the level of
individual features. We propose it as a transparent baseline for CESM classification work that follows, including deep-learning pipelines, under matched patient-grouped evaluation.

\paragraph{Reproducibility}
All code (\textsc{MATLAB} for the feature extraction and classifier pipelines, plus driver scripts for the modality comparison, bootstrap, fusion, classifier comparison, and ResNet-50 frozen-feature analyses) and all per-fold prediction files are available at \url{https://github.com/saraantonijevic/Masked_Mammograms}. The CDD-CESM dataset is publicly available from~\citep{Khaled2021Dataset, Khaled2022ScientificData}.

\bibliographystyle{plainnat}
\bibliography{ref}

\end{document}